\documentclass[twocolumn,superscriptaddress,preprintnumbers,amsmath,amssymb,prl,floatfix]{revtex4-1}

\pdfoutput=1
\usepackage{graphicx} 
\usepackage{dcolumn} 
\usepackage{bm}
\usepackage{color}
\usepackage{hyperref}
\usepackage{graphicx}
\usepackage{amsmath}
\usepackage{setspace}
\usepackage{CJKutf8}

\begin{document}

\preprint{RIKEN-iTHEMS-Report-22}

\title{A Kohn-Sham Scheme Based Neural Network for Nuclear Systems}

\author{Zu-Xing Yang}
\affiliation{RIKEN Nishina Center, Wako 351-0198, Japan}

\author{Xiao-Hua Fan}
\affiliation{School of Physical Science and Technology, Southwest University, Chongqing 400715, China}

\author{Zhi-Pan Li}
\affiliation{School of Physical Science and Technology, Southwest University, Chongqing 400715, China}

\author{Haozhao Liang}
\affiliation{Department of Physics, Graduate School of Science, The University of Tokyo, Tokyo 113-0033, Japan}
\affiliation{RIKEN Interdisciplinary Theoretical and Mathematical Sciences Program, Wako 351-0198, Japan}

\begin{abstract}
A Kohn-Sham scheme based multi-task neural network is elaborated for the supervised learning of nuclear shell evolution. 
The training set is composed of the single-particle wave functions and occupation probabilities of 320 nuclei, calculated by the Skyrme density functional theory. 
It is found that the deduced density distributions, momentum distributions, and charge radii are in good agreements with the benchmarking results for the untrained nuclei. 
In particular, accomplishing shell evolution leads to a remarkable improvement in the extrapolation of nuclear density. 
After a further charge-radius-based calibration, the network evolves a stronger predictive capability.
This opens the possibility to infer correlations among observables by combining experimental data for nuclear complex systems.

\end{abstract}

\maketitle


{\it Introduction}---The last decade has seen the independent machine-learning studies on different nuclear observables to meet the experimental values, such as nuclear masses \cite{Niu2018Phys.Lett.B778.4853, Ma2020Phys.Rev.C101.045204, Athanassopoulos2004Nucl.Phys.A743.222235}, charge radii \cite{Utama2016J.Phys.GNucl.Part.Phys.43.114002, Wu2020Phys.Rev.C102.054323, Dong2022Phys.Rev.C105.014308, Cotextquotesingle2022Phys.Rev.C105.034320}, excited states \cite{Lasseri2020Phys.Rev.Lett.124.162502, Wang2022Phys.Lett.B830.137154},  $\alpha$-decay half-lives \cite{Saxena2021J.Phys.GNucl.Part.Phys.48.055103}, $\beta$-decay half-lives \cite{Niu2019Phys.Rev.C99.064307}, fission yields \cite{Wang2019Phys.Rev.Lett.123.122501, Qiao2021Phys.Rev.C103.034621}, and etc.
Traditionally, these basic nuclear properties are evaluated and discussed using various theoretical approaches with different interactions, amongst which the density functional theory, based on the Hohenberg-Kohn theorem, has been widely recognized due to the accurate and universal calculations. 
On this basis, many branching methods have been developed to cope with various behaviors of nuclei, such as introducing collective Hamiltonian to handle collective vibrations and rotations \cite{Niksic2011Prog.Part.Nucl.Phys.66.519548}, combining random phase approximation methods to describe giant resonances \cite{Liang2022., Paar2007Rep.Progr.Phys.70.R02, RocaMaza2018Prog.Part.Nucl.Phys.101.96176}, and using Woods-Saxon basis expansions to explain halos \cite{Zhou2011J.Phys.Conf.Ser.312.092067, Long2010Phys.Rev.C81.031302}.
Nevertheless, the theoretical high computational cost results in difficulty in finding the universal functional.
Amidst the researches on density functionals, attention has also been paid to the long-standing structure problem of the correlations among different  physical quantities, where the recent breakthroughs have occurred in the physics of electronic systems and condensed matter by combining machine learning techniques.
It is significative that artificial intelligence has reconstructed the Hohenberg-Kohn map of density functional theory \cite{Moreno2020Phys.Rev.Lett.125.076402}, which is the bijection between the local density and the ground-state many-body wave function.
Meanwhile, the self-consistent charge densities calculated with different exchange-correlation functionals were used to make predictions for the correlation, exchange, external, kinetic, and total energies simultaneously by an extensive deep neural network \cite{Ryczko2019Phys.Rev.A100.022512}.

From a fundamental perspective, the effectiveness of density functionals in one respect is a result of the fact that Kohn and Sham introduced explicitly noninteracting particle systems to calculate shell evolution including exchange and correlation effects \cite{Kohn1965Phys.Rev.140.A1133A1138}, which is instrumental in consolidating intrinsic connections of observables.
Such the connections have been effectively exploited in our previous work \cite{Yang2022.}, in which the influences on densities and binding energies were seen with the charge radius residuals between the experimental and theoretical values eliminated by a series of neural networks. 

Inspired by these studies, we are aware that accessing different physical quantities via a single neural network can provide an opportunity to analyze nuclear properties physically and propagate residuals accurately.
In this work, we will construct a Kohn-Sham scheme based multi-task neural network to straightforwardly generate nuclear auxiliary single-particle states for characterizing density functionals, which we will note as Kohn-Sham network (KSN). 



{\it Kohn-Sham network}---To achieve a high-precision neural network, the single-particle wave functions and occupation probabilities are first computed using the Skyrme Hartree-Fock (SHF) approach with SkM* interaction \cite{Bartel1982Nucl.Phys.A386.79100} including the Bardeen-Cooper-Schrieffer (BCS) pairing.
The Kohn-Sham equation reads
\begin{equation}
    (h_{kin} + h_{u} + h_{ls}) \varphi_{i} =\epsilon_{i} \varphi_{i},
\end{equation}
where $h_{kin}$, $h_{u}$, and $h_{ls}$ denote the kinetic, potential, and spin-orbit terms, respectively.
The $\varphi_{i}$ and $\epsilon_{i}$ correspond the single-particle wave function and energy of state $S_i$.
The occupation probability $w_i$ is obtained by solving the BCS equation. 
The spherical symmetry is adopted in the present calculations.
In particular, the conservation of particle number
\begin{equation}
    \label{eq_pc}
    \sum_i w_{i} d_i = A^\tau
\end{equation}
should be obeyed during the reconstruction, where $d_i$ is the degeneracy of state $S_i$ and $A^\tau$ is proton number $Z$ ($\tau=p$) or neutron number $N$ ($\tau=n$).
The challenges occur in how to encapsulate these properties as well as the BCS equation in a neural network.
Facing these difficulties, we note the advantage of taking $\sqrt{w_{i}}\varphi_{i}(r)$ as the predictions, which allows us to consider $\varphi_{i}$ as normalized and exempts us from modelling nonlinear changes in the pairing effect. 
In this case, one only needs to be rigorous about particle number conservation and orthogonality among different single-particle states.

\begin{figure}
\includegraphics[width=8.5 cm]{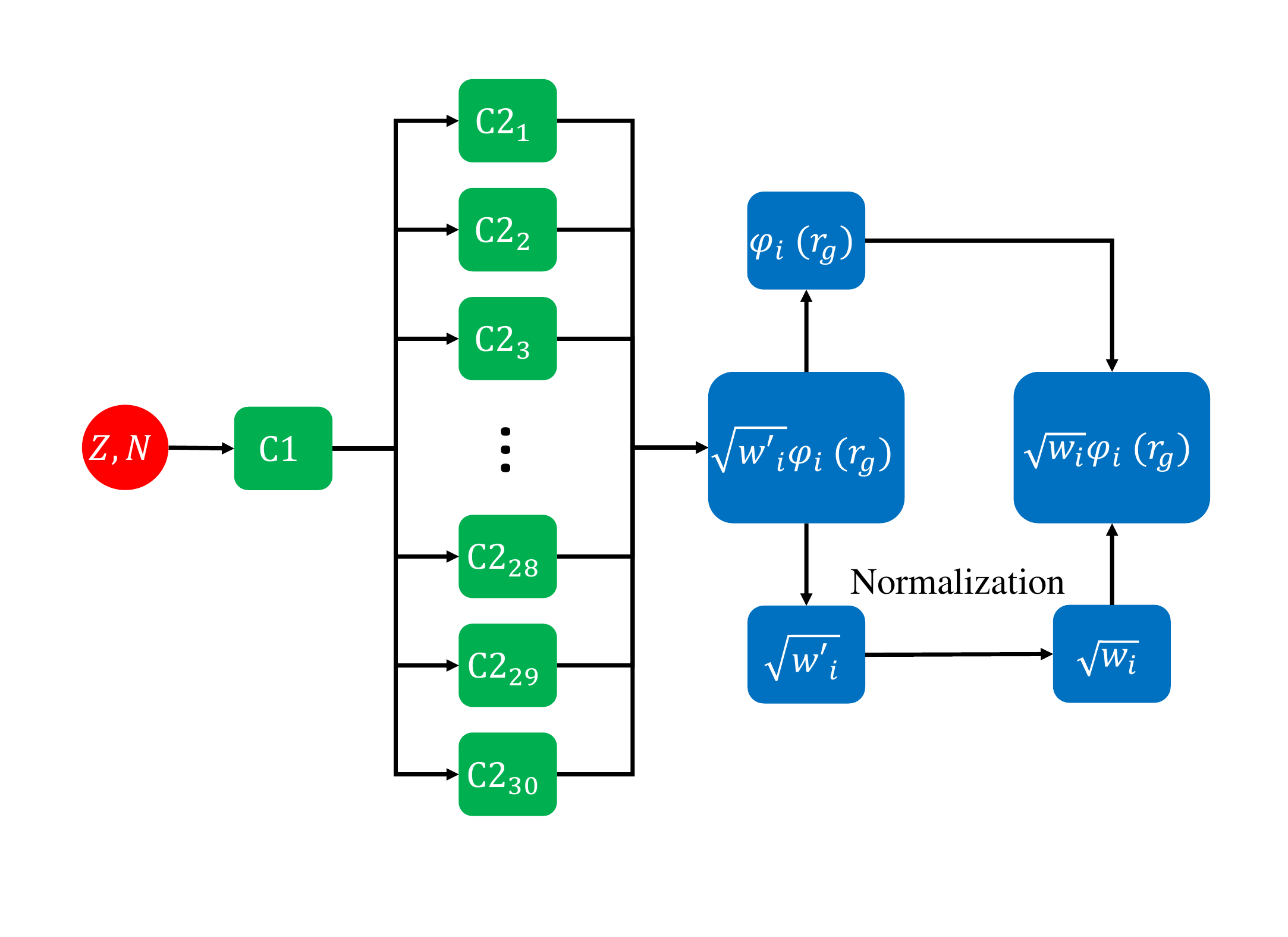}
\caption{\label{fig_structure} (Color online) Schematic diagram of the structure of Kohn-Sham network (KSN).}
\end{figure}
The structure of KSN designed based on the above discussions is shown in Fig.~\ref{fig_structure}.
The input of the network is $\mathbf{x} = \{Z,N\}$ and the output is a matrix with the element $\sqrt{w_{i}}\varphi_{i}(r_g)$~$(i=1,2,...,30)$ with $r_g = 0.1\, \mathrm{fm} \times g$~$(g=0,1,...,149)$.
In detail, a multi-task fully connected (FC) neural network is selected, where the FC layer is written as
\begin{equation}
a_{k}^{(l+1)} =g\left(b_{k}^{(l)}+\sum_{j=1}^{C_{l}} w_{k j}^{(l)} a_{j}^{(l)}\right) ,
\end{equation}
with $a_{j}^{(l)}$ and $a_{k}^{(l+1)}$ denoting the input and output of the $l$-th layer, $w_{k j}^{(l)}$ and $b_{k}^{(l)}$ being the trainable weight and bias, and ${C_{l}}$ corresponding the number of neurons.
The nonlinear activation functions $g(x) =\mathrm{ReLU}(x)$ and $g(x) = \mathrm{tanh}(x)$ are employed for different layers.
Firstly, $\{Z,N\}$ is fed into a five-layer FC neural network trunk cell ($\mathrm{C1}$) to produce latent features.
Subsequently, the latent features are entered into each of the 30 FC neural network branch cells with the same structure ($\mathrm{C2}$), corresponding to the 30 single-particle states, for which the typical shell ordering according to the nuclear oscillator shell model is taken for the state $S_i$ listing in Supplemental Materials.
Obviously, the latent features contain the associations among different shells.
This process is recorded as
\begin{equation}
\sqrt{w'_i} \varphi_i(r) = \mathrm{C2}_i(\mathrm{C1}(\mathbf{x})).
\label{eq_5}
\end{equation}
In practice, we train on Eq.~(\ref{eq_5}) as the first step with the objective function being
\begin{equation}
\label{eq_loss1}
\mathrm{Loss}_1 = \sum_i d_i \int (\sqrt{w'_i} \varphi_i(r)- y_i)^2 dr \times 1\, \mathrm{fm}^{2},
\end{equation}
where $y_i = \sqrt{w_{i,tar}} \varphi_{i,tar}(r)$ is the target gained by SHF+BCS and degeneracy $d_i$ weights each shell.
The feasibility and generalizability of this training has been discussed in Ref.~\cite{Yang2021Phys.Lett.B823.136650}.
After 1500 epochs (about 20 GPU minutes), the trainable parameters of the pre-trained model are determined initially.

Considering the conservation of particle number (Eq.~(\ref{eq_pc})) and the normalized single-particle wave function, the excess nucleon number $\Delta$, as a small value due to the activation function $\mathrm{tanh}(x)$ in C2 and the pre-trained model, is superimposed on the outermost shell according to typical shell ordering, which is referred to as normalization denoted by
\begin{equation}
\sqrt{w_i} \varphi_i(r) = \hat{N}(\sqrt{w'_i} \varphi_i(r)).
\label{eq_nOR}
\end{equation}
In implementation, the normalization operator and the corresponding back-propagation are achieved with the help of {\bf Autograd} in Pytorch \cite{Ketkar2017.195208}.
The pre-trained parameters are loaded into the calibrated network as the second step for fine-tuning, i.e., another 500 epochs (about 15 GPU minutes) are preformed under the objective function
\begin{equation}
\label{eq_loss2}
\begin{aligned}
\mathrm{Loss}_2 =& \sum_i d_i \int (\sqrt{w_i} \varphi_i(r)- y_i)^2 dr \times 1\, \mathrm{fm}^{2} \\
&+ (\mu \Delta)^2,
\end{aligned}
\end{equation}
where the $(\mu \Delta)^2$ punishes mass number deviation with $\mu=0.05$ balancing the right two terms in magnitude and the factor $1\, \mathrm{fm}^{2}$ makes $\mathrm{Loss}_1$ and $\mathrm{Loss}_2$ dimensionless.

Actually, the normalization substantially complicates the gradient, which would prolong the back-propagation and even cause gradient vanishing.
To feasibly operationalize this training, we have to decompose the whole process into the two parts above and vary the learning rate according to the loss value of stochastic gradient descent, where Adaptive Momentum Estimation (Adam) \cite{Kingma2015.} is taken as the optimizer.
The carefully modulated hyperparameters sets of the FC neural network cells, the normalization details, and the varying learning rate are listed in Supplemental Materials.


\begin{figure}
\includegraphics[width=9 cm]{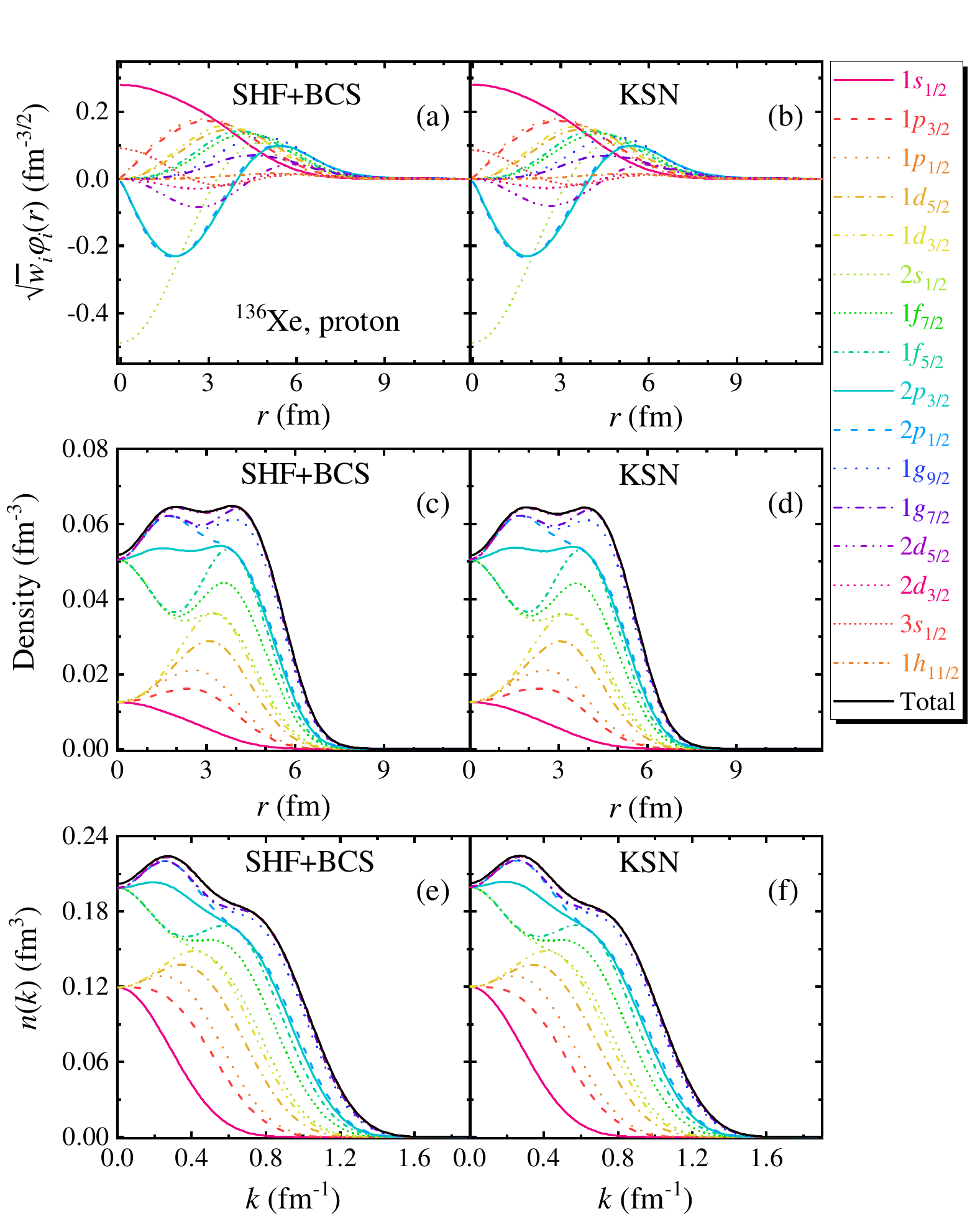}
\caption{\label{fig_predict} (Color online)
Upper panels: The $\sqrt{w_i} \varphi_i(r)$ in coordinate space for each single-particle orbital for the untrained nucleus $^{136}$Xe.
Middle and lower panels: The corresponding proton densities and momentum distributions as well as the contributions from each orbital.
The predictions by KSN are compared with the calculations by SHF+BCS.}
\end{figure}

{\it Results}---
In application, we select only a few nuclei (320 nuclei) as the training set, which are $\sim 10\%$ of the nuclei discovered so far, but the network shows a remarkable generalization capability.
Taken the untrained nucleus $^{136}$Xe as an example, the properties of proton single-particle states, densities, and momenta are exhibited in Fig.~\ref{fig_predict}.
Panels (a) and (b) show the $\sqrt{w_{i}} \varphi_{i}(r)$ in coordinate space  for each single-particle orbital calculated by SHF+BCS and predicted by KSN, respectively.
It is obvious that the intrinsic orbital information on theoretical calculations and neural network predictions is indistinguishable by eyes.
As a more direct test, we present the density distributions deduced by (c) SHF+BCS and (d) KSN.
Each curve is obtained by superimposing orderedly single-particle densities from the state $i=1$ to a corresponding state $I$,
\begin{equation}
\rho_I(r) = \sum_{i=1}^I \frac{d_i(\sqrt{w_{i}} \varphi_{i}(r))^2 }{4\pi}.  
\end{equation}
The outermost curve indicates the total density of the nucleus and the area between states $I-1$ and $I$ represents the contribution of state $S_I$ to the total density.
Similar to the $\sqrt{w_{i}} \varphi_{i}(r)$, the total densities and the contributions of each state from SHF+BCS and KSN cannot be distinguished either.
In addition, the single-particle wave functions in momentum space derived by Fourier transform are utilized to calculate the momentum distributions and the contributions of each state, displayed in (e) and (f) with the formula
\begin{equation}
n_I(k) = \sum_{i=1}^I \frac{d_i(\sqrt{w_{i}} \varphi_{i}(k))^2 }{4\pi A^\tau},  
\end{equation}
which yields the same conclusion as density.
It is clear that the other currents and densities, such as spin-orbit current and kinetic energy density, would have the same quality of outcomes.
The plots show that the network can accurately predict all of the wave function components, without displaying significant deviations for magnitude, gradient and integration.
Achieving a high degree of accuracy on examples absent from the training set is an indication that the network has not been overfitted, and indeed has captured the underlying physical connotation of Kohn-Sham equation and the auxiliary noninteracting particle systems.

\begin{figure}[bt]
\includegraphics[width=8.5 cm]{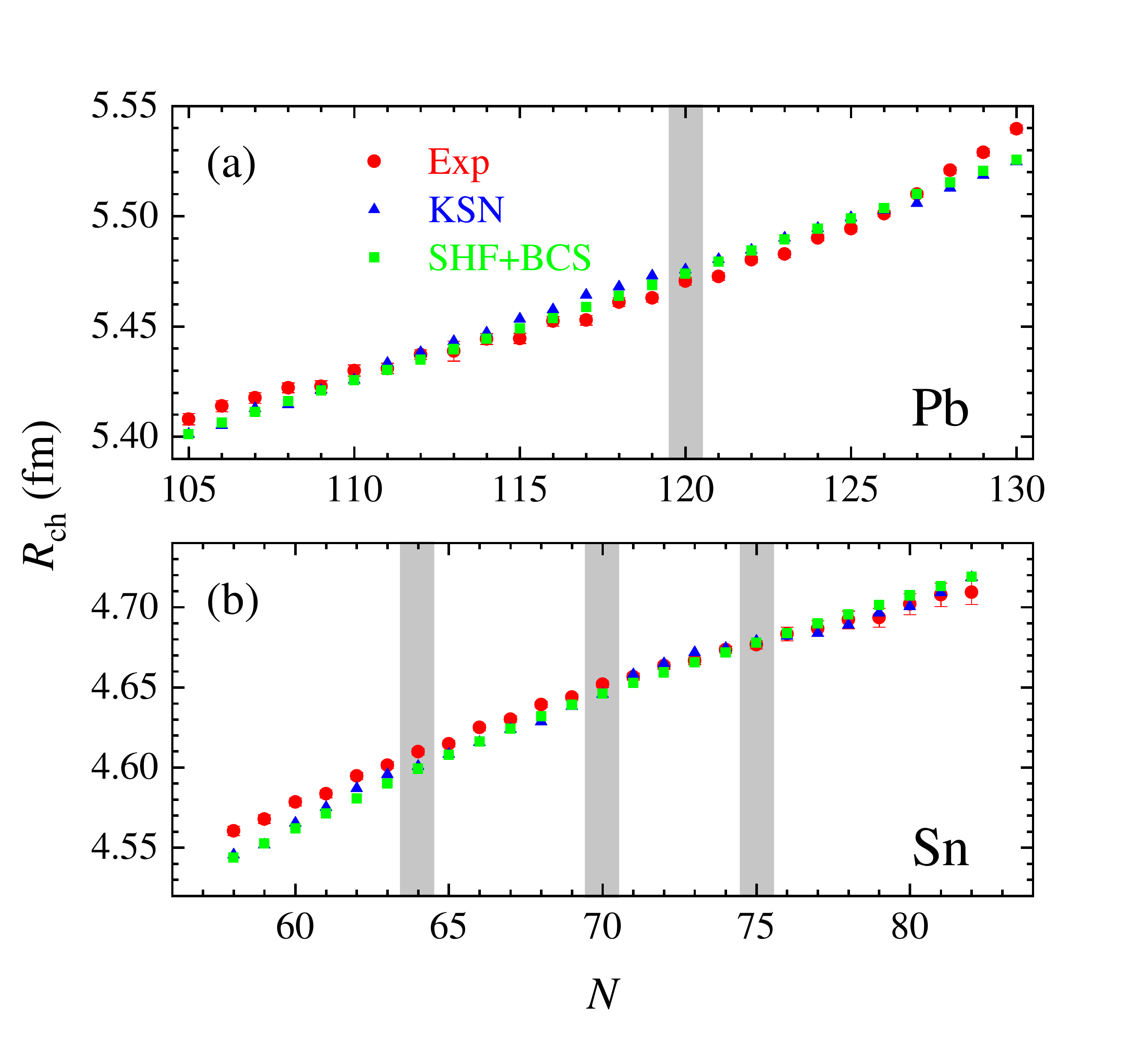}
\caption{\label{fig_Rch} (Color online) Charge radii predicted by KSN for (a) Pb and (b) Sn isotopes, where the training regions are indicated by shadows.
The SHF+BCS results and experimental data (Exp) are also shown for comparison.}
\end{figure}

The charge radius $R_\text{ch}$ is one of the experimental important observables, which is phenomenologically formulated as \cite{Sugahara1994Nucl.Phys.A579.557572}
\begin{equation}
R_\text{ch}^2 = R_p^2 + (0.862\, \mathrm{fm})^2 - (0.336\, \mathrm{fm})^2 \frac{N}{Z}, 
\end{equation}
where $R_p$ indicates radius of the proton distribution, the second and third terms are due to the proton size and neutron size, respectively.
Note that there are also contributions by the spin-orbit effects to the charge radius \cite{Reinhard1991.2850}, while such contributions are relatively small for the nuclei with $Z > 40$. 
The calculations for Pb and Sn isotopes are shown in Fig.~\ref{fig_Rch} (a) and (b), respectively, where the shadows mark the nuclei in the training set.
Most of the radii calculated by SHF+BCS deviate from the experiment by less than 0.01 fm, which demonstrates that the theoretical calculations are quite accurate.
The KSN reaches almost the same accuracy as the theoretical calculations by training only four nuclei within the display region.
A step further than traditional computing is that neural networks allow the direct involvement of experiment \cite{Yang2022.}, i.e., single-particle states have the prospect of being calibrated exactly by a large amount of experimental charge radius and binding energy data, which will facilitate the realization of realistic non-parametric Hohenberg-Kohn map for nuclear complex systems.

\begin{figure}[tb]
\includegraphics[width=9 cm]{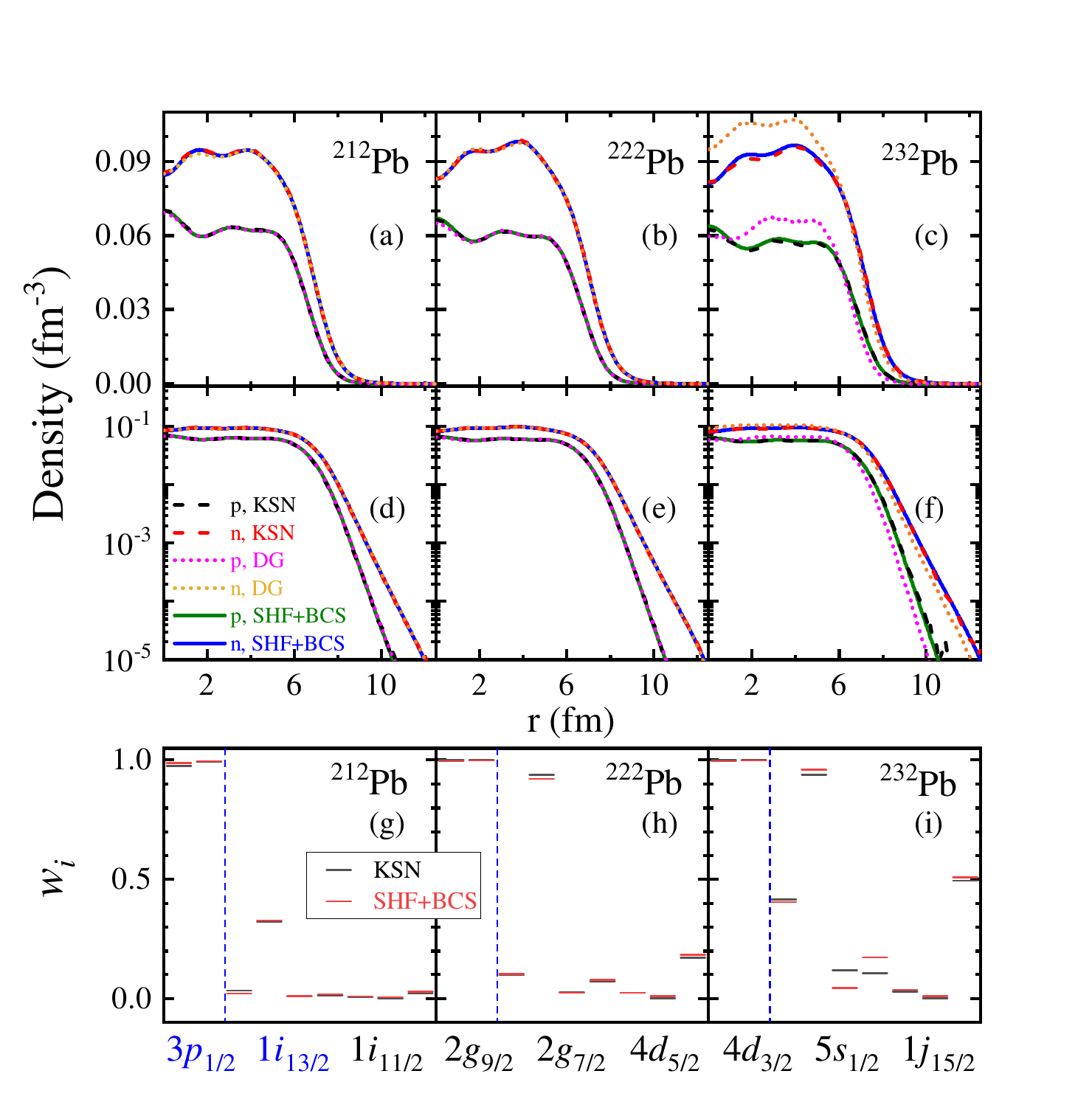}
\caption{\label{fig_Extrap} (Color online) Neutron and proton densities (shown in linear and logarithmic scales) as well as the neutron occupation probabilities $w_i$ in valence space for the extrapolated nuclei $^{212}$Pb, $^{222}$Pb, and $^{232}$Pb.
See the text for details.}
\end{figure}

As a further in-depth analysis, the superiority of the KSN extrapolation is illustrated in Fig.~\ref{fig_Extrap} by taking the Pb isotopes as examples, where the most neutron-rich trained nucleus is $^{202}$Pb, and we extrapolate to $^{212}$Pb, $^{222}$Pb, and $^{232}$Pb.
It is evident from Fig.~\ref{fig_Extrap}(a)-(f) that the KSN reproduces the theoretical density distributions almost identically both in the linear and logarithmic scales, even in the extrapolation of an additional 30 neutrons.
In contrast, the extrapolation of density generator (DG) with simply learning the densities in Ref.~\cite{Yang2021Phys.Lett.B823.136650} appears an obvious deviation in $^{232}$Pb.
Neural networks are often considered as an excellent interpolator, but an unreliable extrapolator.
Our conclusions demonstrate that the extrapolation performance of networks naturally improves with a reduced degree of freedom as more physical natures are learned and more constraints are imposed.
The present extrapolation is still far from the drip-line \cite{Xia2018At.DataNucl.DataTables121122.1215}, therefore we select randomly 2400 nuclei found in laboratories to extrapolate farther for Pb, which are shown in Supplemental Materials.
More trained nuclei make the predictions more robust, while the training time increases linearly.

To make sense of the subtle discrepancies, the neutron occupation probabilities in valence space are shown in Fig.~\ref{fig_Extrap}(g)-(i), where the short solid lines correspond to the occupation probabilities of single-particle states ($3p_{1/2}$, $1i_{13/2}$, $1i_{11/2}$, $2g_{9/2}$, $2g_{7/2}$, $4d_{5/2}$, $4d_{3/2}$, $5s_{1/2}$, and $1j_{15/2}$) in order from left to right, and the blue vertical line separate the valence space from the rest.
The occupation probabilities for protons and deep-bound-state neutrons are neglected due to the completely filled shells.
For $^{212}$Pb and $^{222}$Pb, KSN succeeds in describing the occupation probabilities of single-particle states for valence nucleons, both for the ordering and the magnitude.
In case of $^{232}$Pb, the states $2g_{7/2}$, $4d_{5/2}$ deviate significantly from the theoretical calculations, which originates from the increasing excess nucleon number $\Delta$ with extrapolation.
The excess nucleon number $\Delta = 0.64$ is superimposed on $2g_{7/2}$ in accordance with the normalization and the given typical shell ordering.
Other than that, it still accurately predicts other states, which ensures that the overall deviation of the density is not too large.
We can conclude that shell evolution from the single-particle perspective is indeed encompassed by the network, which is the foundation for the strong extrapolation performance.

Additionally, the orthogonality of the generated single-particle wave functions is investigated for $^{212,222,232}$Pb.
By definition, the angular orthogonality is provided by spinor spherical harmonics, thus only the radial one needs to be discussed.
We specify
\begin{equation}
{O}_{ik} \equiv \int \sqrt{w_iw_k} \varphi_{i}(r) \varphi_{k}(r) r^2 dr,
\end{equation}
where ${O}_{ik}$ should be zero in principle as long as $i$ and $k$ belong to $P = \{i,k | {n}_i \neq {n}_k,~{l}_i = {l}_k,~{j}_i = {j}_k \}$.
As an inspection, the root-mean-square of ${O}_{ik}$ is calculated for each nucleus,
\begin{equation}
\mathcal{O} = \sqrt{\frac{1}{\mathrm{card}(P)}\sum_{i,k \in P} {O}_{ik}^2},
\end{equation}
where $\mathrm{card}(P)$ is the aggregate of set $P$.
Calculations show $\mathcal{O}(^{212}\mathrm{Pb})$ = 0.0021, $\mathcal{O}(^{222}\mathrm{Pb})$ = 0.0022, and $\mathcal{O}(^{232}\mathrm{Pb})$ = 0.0036. 
In this regard, the results are acceptable to a large extent, which lead to not imposing relevant penalties on the network.
Affirmatively, the risk that the orthogonality deviation increases progressively with extrapolation deserves attention as a result of the possible deteriorating description of the density.

\begin{figure}[tb]
\includegraphics[width=9 cm]{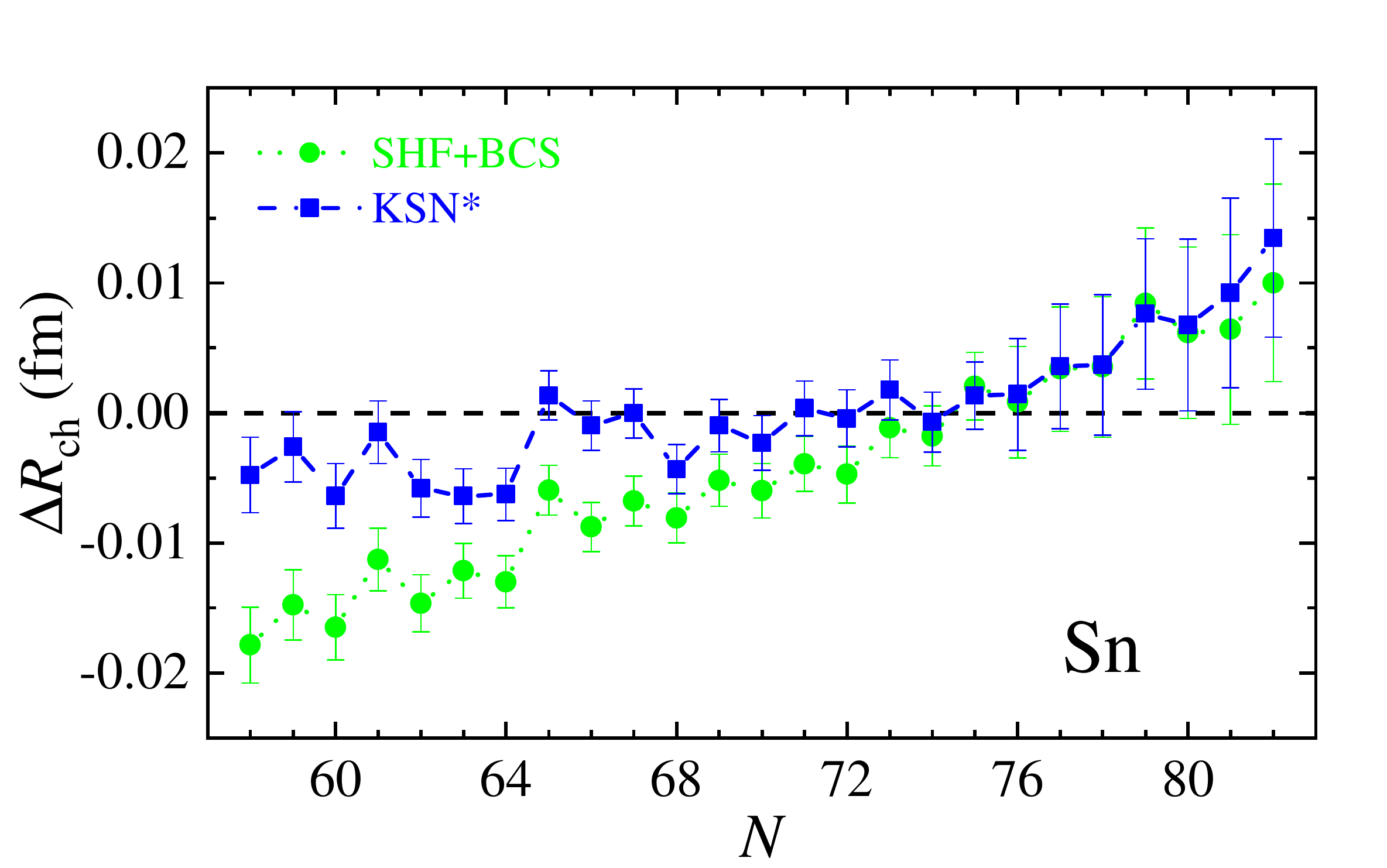}
\caption{\label{fig_CaliSn} (Color online) Deviation in charge radii ($\Delta R_\text{ch}$) between the calibrated-KSN (KSN*) predictions and experimental data for the untrained Sn isotopes, where the error bars representing the experimental errors. 
The deviation between SHF+BCS results and experimental data are also shown for comparison.}
\end{figure}

{\it Calibration for charge radius}---To go beyond the theoretical calculations, we include a calibration of the experimental charge radius in KSN by further considering the objective function as
\begin{equation}
\label{eq_loss3}
\mathrm{Loss}_3 = \mathrm{Loss}_2 + (\Delta R_\text{ch})^2 \times 0.1\, \mathrm{fm}^{-2},
\end{equation}
where $\Delta R_\text{ch}$ is the deviation in charge radii between the calibrated-KSN (noted as KSN*) predictions and experimental data. 
In this case, the objective function will aim to make the smallest possible correction to the theoretical calculations.
More than 600 nuclides with $Z>40$ are taken as calibration, whereas the Sn isotopes are completely excluded in order to examine the calibration, the results of which are presented in Fig.~\ref{fig_CaliSn}.
Evidently, around the stable Sn isotopes with more accurate experimental data, the prediction gains a significant improvement.
Meanwhile, it is noticed that heavy Sn isotopes still exhibit a bias toward the theoretical results, which can be attributed to two factors: 
1. The network fails to capture the necessary calibration information from the training set, which may be caused by the experimental precision; 
2. the theoretical calculations still carry considerable weight in calibration. 
We conclude that the calibrated network has a stronger prediction capability even for these untrained nuclei, although it still partly relies on the theoretical calculations. 
In other words, the theoretical calculations and KSN-based calibrations are complementary to each other.

The neural network method benefiting from current nuclear structure models has ample potential to tackle some nuclear physics problems, which can in turn be employed to improve current nuclear structure models. 
The calibrated single-particle wave functions allow for the natural derivation of new spin-orbit and kinetic densities. 
Then, new nuclear effective interactions could be constructed with a neural network framework, where the back-propagation principle can be leveraged to incorporate experimental binding energies into the functionals efficiently. 
These functionals can be transferred to current nuclear structure models, which may lead to better descriptions and predictions of nuclear properties.



{\it Conclusion}---A novel supervised deep multi-task learning on the nuclear ground-state shell evolution from the Kohn-Sham single-particle perspective has been successfully constructed in a 1D lattice.
The carefully designed KSN and its training process have taken the conservation of particle number and the orthonormality of single-particle wave function into account, whereby the physics features embedded in nuclear density functional can be preserved to a large extent.
The success of decomposing the network so as to train it step by step is illuminating for solving the problem of barren plateaus in large complex networks.

With only 320 nuclei trained, KSN generates proton wave function components, density, and momentum for the untrained nucleus $^{136}$Xe that are indistinguishable from the SHF+BCS results, besides achieving a high accuracy description for charge radii of Pb and Sn isotopes.
Succeeding basically in portraying the densities and occupation probabilities of valence nucleons for $^{212,222,232}$Pb serves as a proof of reliable extrapolation performance to KSN, which significantly outperforms the one with simply learning the densities, demonstrating that extrapolation of networks naturally improves amid paying attention to shell evolution.
Our conclusions strengthen evidence that machine learning tools provide a suitable framework to represent density functionals, analyze physical correlation effects, and calibrate theoretical calculations with experiment data.
After a further charge-radius-based calibration, the network indeed evolves a stronger predictive capability, which is confirmed by the untrained nuclei, e.g., the Sn isotopes.
Furthermore, the calibrated single-particle wave functions and densities would facilitate the researches on finding general non-parametric functionals, which would improve nuclear property descriptions and decrease computational costs.

\begin{acknowledgements}
{\it Acknowledgements}---This work is supported by the National Natural Science Foundation of China under Grants No.~12005175,
the Fundamental Research Funds for the Central Universities under Grant No.~SWU119076,
the JSPS Grant-in-Aid for Early-Career Scientists under Grant No.~18K13549,
the JSPS Grant-in-Aid for Scientific Research (S) under Grant No.~20H05648.
This work is also partially supported by the RIKEN Pioneering Project: Evolution of Matter in the Universe.
\end{acknowledgements}

\bibliographystyle{apsrev4-1}
\bibliography{Ref}

\begin{thebibliography}{30}%
\makeatletter
\providecommand \@ifxundefined [1]{%
 \@ifx{#1\undefined}
}%
\providecommand \@ifnum [1]{%
 \ifnum #1\expandafter \@firstoftwo
 \else \expandafter \@secondoftwo
 \fi
}%
\providecommand \@ifx [1]{%
 \ifx #1\expandafter \@firstoftwo
 \else \expandafter \@secondoftwo
 \fi
}%
\providecommand \natexlab [1]{#1}%
\providecommand \enquote  [1]{``#1''}%
\providecommand \bibnamefont  [1]{#1}%
\providecommand \bibfnamefont [1]{#1}%
\providecommand \citenamefont [1]{#1}%
\providecommand \href@noop [0]{\@secondoftwo}%
\providecommand \href [0]{\begingroup \@sanitize@url \@href}%
\providecommand \@href[1]{\@@startlink{#1}\@@href}%
\providecommand \@@href[1]{\endgroup#1\@@endlink}%
\providecommand \@sanitize@url [0]{\catcode `\\12\catcode `\$12\catcode
  `\&12\catcode `\#12\catcode `\^12\catcode `\_12\catcode `\%12\relax}%
\providecommand \@@startlink[1]{}%
\providecommand \@@endlink[0]{}%
\providecommand \url  [0]{\begingroup\@sanitize@url \@url }%
\providecommand \@url [1]{\endgroup\@href {#1}{\urlprefix }}%
\providecommand \urlprefix  [0]{URL }%
\providecommand \Eprint [0]{\href }%
\providecommand \doibase [0]{http://dx.doi.org/}%
\providecommand \selectlanguage [0]{\@gobble}%
\providecommand \bibinfo  [0]{\@secondoftwo}%
\providecommand \bibfield  [0]{\@secondoftwo}%
\providecommand \translation [1]{[#1]}%
\providecommand \BibitemOpen [0]{}%
\providecommand \bibitemStop [0]{}%
\providecommand \bibitemNoStop [0]{.\EOS\space}%
\providecommand \EOS [0]{\spacefactor3000\relax}%
\providecommand \BibitemShut  [1]{\csname bibitem#1\endcsname}%
\let\auto@bib@innerbib\@empty
\bibitem [{\citenamefont {Niu}\ and\ \citenamefont
  {Liang}(2018)}]{Niu2018Phys.Lett.B778.4853}%
  \BibitemOpen
  \bibfield  {author} {\bibinfo {author} {\bibfnamefont {Z.}~\bibnamefont
  {Niu}}\ and\ \bibinfo {author} {\bibfnamefont {H.}~\bibnamefont {Liang}},\
  }\href {\doibase 10.1016/j.physletb.2018.01.002} {\bibfield  {journal}
  {\bibinfo  {journal} {Phys. Lett. B}\ }\textbf {\bibinfo {volume} {778}},\
  \bibinfo {pages} {48} (\bibinfo {year} {2018})}\BibitemShut {NoStop}%
\bibitem [{\citenamefont {Ma}\ \emph {et~al.}(2020)\citenamefont {Ma},
  \citenamefont {Bao}, \citenamefont {Niu}, \citenamefont {Zhao},\ and\
  \citenamefont {Arima}}]{Ma2020Phys.Rev.C101.045204}%
  \BibitemOpen
  \bibfield  {author} {\bibinfo {author} {\bibfnamefont {C.}~\bibnamefont
  {Ma}}, \bibinfo {author} {\bibfnamefont {M.}~\bibnamefont {Bao}}, \bibinfo
  {author} {\bibfnamefont {Z.~M.}\ \bibnamefont {Niu}}, \bibinfo {author}
  {\bibfnamefont {Y.~M.}\ \bibnamefont {Zhao}}, \ and\ \bibinfo {author}
  {\bibfnamefont {A.}~\bibnamefont {Arima}},\ }\href {\doibase
  10.1103/physrevc.101.045204} {\bibfield  {journal} {\bibinfo  {journal}
  {Phys. Rev. C}\ }\textbf {\bibinfo {volume} {101}},\ \bibinfo {pages}
  {045204} (\bibinfo {year} {2020})}\BibitemShut {NoStop}%
\bibitem [{\citenamefont {Athanassopoulos}\ \emph {et~al.}(2004)\citenamefont
  {Athanassopoulos}, \citenamefont {Mavrommatis}, \citenamefont {Gernoth},\
  and\ \citenamefont {Clark}}]{Athanassopoulos2004Nucl.Phys.A743.222235}%
  \BibitemOpen
  \bibfield  {author} {\bibinfo {author} {\bibfnamefont {S.}~\bibnamefont
  {Athanassopoulos}}, \bibinfo {author} {\bibfnamefont {E.}~\bibnamefont
  {Mavrommatis}}, \bibinfo {author} {\bibfnamefont {K.}~\bibnamefont
  {Gernoth}}, \ and\ \bibinfo {author} {\bibfnamefont {J.}~\bibnamefont
  {Clark}},\ }\href {\doibase 10.1016/j.nuclphysa.2004.08.006} {\bibfield
  {journal} {\bibinfo  {journal} {Nucl. Phys. A}\ }\textbf {\bibinfo {volume}
  {743}},\ \bibinfo {pages} {222} (\bibinfo {year} {2004})}\BibitemShut
  {NoStop}%
\bibitem [{\citenamefont {Utama}\ \emph {et~al.}(2016)\citenamefont {Utama},
  \citenamefont {Chen},\ and\ \citenamefont
  {Piekarewicz}}]{Utama2016J.Phys.GNucl.Part.Phys.43.114002}%
  \BibitemOpen
  \bibfield  {author} {\bibinfo {author} {\bibfnamefont {R.}~\bibnamefont
  {Utama}}, \bibinfo {author} {\bibfnamefont {W.-C.}\ \bibnamefont {Chen}}, \
  and\ \bibinfo {author} {\bibfnamefont {J.}~\bibnamefont {Piekarewicz}},\
  }\href {\doibase 10.1088/0954-3899/43/11/114002} {\bibfield  {journal}
  {\bibinfo  {journal} {J. Phys. G: Nucl. Part. Phys.}\ }\textbf {\bibinfo
  {volume} {43}},\ \bibinfo {pages} {114002} (\bibinfo {year}
  {2016})}\BibitemShut {NoStop}%
\bibitem [{\citenamefont {Wu}\ \emph {et~al.}(2020)\citenamefont {Wu},
  \citenamefont {Bai}, \citenamefont {Sagawa},\ and\ \citenamefont
  {Zhang}}]{Wu2020Phys.Rev.C102.054323}%
  \BibitemOpen
  \bibfield  {author} {\bibinfo {author} {\bibfnamefont {D.}~\bibnamefont
  {Wu}}, \bibinfo {author} {\bibfnamefont {C.~L.}\ \bibnamefont {Bai}},
  \bibinfo {author} {\bibfnamefont {H.}~\bibnamefont {Sagawa}}, \ and\ \bibinfo
  {author} {\bibfnamefont {H.~Q.}\ \bibnamefont {Zhang}},\ }\href {\doibase
  10.1103/physrevc.102.054323} {\bibfield  {journal} {\bibinfo  {journal}
  {Phys. Rev. C}\ }\textbf {\bibinfo {volume} {102}},\ \bibinfo {pages}
  {054323} (\bibinfo {year} {2020})}\BibitemShut {NoStop}%
\bibitem [{\citenamefont {Dong}\ \emph {et~al.}(2022)\citenamefont {Dong},
  \citenamefont {An}, \citenamefont {Lu},\ and\ \citenamefont
  {Geng}}]{Dong2022Phys.Rev.C105.014308}%
  \BibitemOpen
  \bibfield  {author} {\bibinfo {author} {\bibfnamefont {X.-X.}\ \bibnamefont
  {Dong}}, \bibinfo {author} {\bibfnamefont {R.}~\bibnamefont {An}}, \bibinfo
  {author} {\bibfnamefont {J.-X.}\ \bibnamefont {Lu}}, \ and\ \bibinfo {author}
  {\bibfnamefont {L.-S.}\ \bibnamefont {Geng}},\ }\href {\doibase
  10.1103/physrevc.105.014308} {\bibfield  {journal} {\bibinfo  {journal}
  {Phys. Rev. C}\ }\textbf {\bibinfo {volume} {105}},\ \bibinfo {pages}
  {014308} (\bibinfo {year} {2022})}\BibitemShut {NoStop}%
\bibitem [{\citenamefont {Co{\textquotesingle}}\ \emph
  {et~al.}(2022)\citenamefont {Co{\textquotesingle}}, \citenamefont
  {Anguiano},\ and\ \citenamefont
  {Lallena}}]{Cotextquotesingle2022Phys.Rev.C105.034320}%
  \BibitemOpen
  \bibfield  {author} {\bibinfo {author} {\bibfnamefont {G.}~\bibnamefont
  {Co{\textquotesingle}}}, \bibinfo {author} {\bibfnamefont {M.}~\bibnamefont
  {Anguiano}}, \ and\ \bibinfo {author} {\bibfnamefont {A.~M.}\ \bibnamefont
  {Lallena}},\ }\href {\doibase 10.1103/physrevc.105.034320} {\bibfield
  {journal} {\bibinfo  {journal} {Phys. Rev. C}\ }\textbf {\bibinfo {volume}
  {105}},\ \bibinfo {pages} {034320} (\bibinfo {year} {2022})}\BibitemShut
  {NoStop}%
\bibitem [{\citenamefont {Lasseri}\ \emph {et~al.}(2020)\citenamefont
  {Lasseri}, \citenamefont {Regnier}, \citenamefont {Ebran},\ and\
  \citenamefont {Penon}}]{Lasseri2020Phys.Rev.Lett.124.162502}%
  \BibitemOpen
  \bibfield  {author} {\bibinfo {author} {\bibfnamefont {R.-D.}\ \bibnamefont
  {Lasseri}}, \bibinfo {author} {\bibfnamefont {D.}~\bibnamefont {Regnier}},
  \bibinfo {author} {\bibfnamefont {J.-P.}\ \bibnamefont {Ebran}}, \ and\
  \bibinfo {author} {\bibfnamefont {A.}~\bibnamefont {Penon}},\ }\href
  {\doibase 10.1103/physrevlett.124.162502} {\bibfield  {journal} {\bibinfo
  {journal} {Phys. Rev. Lett.}\ }\textbf {\bibinfo {volume} {124}},\ \bibinfo
  {pages} {162502} (\bibinfo {year} {2020})}\BibitemShut {NoStop}%
\bibitem [{\citenamefont {Wang}\ \emph {et~al.}(2022)\citenamefont {Wang},
  \citenamefont {Zhang}, \citenamefont {Niu},\ and\ \citenamefont
  {Li}}]{Wang2022Phys.Lett.B830.137154}%
  \BibitemOpen
  \bibfield  {author} {\bibinfo {author} {\bibfnamefont {Y.}~\bibnamefont
  {Wang}}, \bibinfo {author} {\bibfnamefont {X.}~\bibnamefont {Zhang}},
  \bibinfo {author} {\bibfnamefont {Z.}~\bibnamefont {Niu}}, \ and\ \bibinfo
  {author} {\bibfnamefont {Z.}~\bibnamefont {Li}},\ }\href {\doibase
  10.1016/j.physletb.2022.137154} {\bibfield  {journal} {\bibinfo  {journal}
  {Phys. Lett. B}\ }\textbf {\bibinfo {volume} {830}},\ \bibinfo {pages}
  {137154} (\bibinfo {year} {2022})}\BibitemShut {NoStop}%
\bibitem [{\citenamefont {Saxena}\ \emph {et~al.}(2021)\citenamefont {Saxena},
  \citenamefont {Sharma},\ and\ \citenamefont
  {Saxena}}]{Saxena2021J.Phys.GNucl.Part.Phys.48.055103}%
  \BibitemOpen
  \bibfield  {author} {\bibinfo {author} {\bibfnamefont {G.}~\bibnamefont
  {Saxena}}, \bibinfo {author} {\bibfnamefont {P.~K.}\ \bibnamefont {Sharma}},
  \ and\ \bibinfo {author} {\bibfnamefont {P.}~\bibnamefont {Saxena}},\ }\href
  {\doibase 10.1088/1361-6471/abcd1c} {\bibfield  {journal} {\bibinfo
  {journal} {J. Phys. G: Nucl. Part. Phys.}\ }\textbf {\bibinfo {volume}
  {48}},\ \bibinfo {pages} {055103} (\bibinfo {year} {2021})}\BibitemShut
  {NoStop}%
\bibitem [{\citenamefont {Niu}\ \emph {et~al.}(2019)\citenamefont {Niu},
  \citenamefont {Liang}, \citenamefont {Sun}, \citenamefont {Long},\ and\
  \citenamefont {Niu}}]{Niu2019Phys.Rev.C99.064307}%
  \BibitemOpen
  \bibfield  {author} {\bibinfo {author} {\bibfnamefont {Z.~M.}\ \bibnamefont
  {Niu}}, \bibinfo {author} {\bibfnamefont {H.~Z.}\ \bibnamefont {Liang}},
  \bibinfo {author} {\bibfnamefont {B.~H.}\ \bibnamefont {Sun}}, \bibinfo
  {author} {\bibfnamefont {W.~H.}\ \bibnamefont {Long}}, \ and\ \bibinfo
  {author} {\bibfnamefont {Y.~F.}\ \bibnamefont {Niu}},\ }\href {\doibase
  10.1103/physrevc.99.064307} {\bibfield  {journal} {\bibinfo  {journal} {Phys.
  Rev. C}\ }\textbf {\bibinfo {volume} {99}},\ \bibinfo {pages} {064307}
  (\bibinfo {year} {2019})}\BibitemShut {NoStop}%
\bibitem [{\citenamefont {Wang}\ \emph {et~al.}(2019)\citenamefont {Wang},
  \citenamefont {Pei}, \citenamefont {Liu},\ and\ \citenamefont
  {Qiang}}]{Wang2019Phys.Rev.Lett.123.122501}%
  \BibitemOpen
  \bibfield  {author} {\bibinfo {author} {\bibfnamefont {Z.-A.}\ \bibnamefont
  {Wang}}, \bibinfo {author} {\bibfnamefont {J.}~\bibnamefont {Pei}}, \bibinfo
  {author} {\bibfnamefont {Y.}~\bibnamefont {Liu}}, \ and\ \bibinfo {author}
  {\bibfnamefont {Y.}~\bibnamefont {Qiang}},\ }\href {\doibase
  10.1103/physrevlett.123.122501} {\bibfield  {journal} {\bibinfo  {journal}
  {Phys. Rev. Lett.}\ }\textbf {\bibinfo {volume} {123}},\ \bibinfo {pages}
  {122501} (\bibinfo {year} {2019})}\BibitemShut {NoStop}%
\bibitem [{\citenamefont {Qiao}\ \emph {et~al.}(2021)\citenamefont {Qiao},
  \citenamefont {Pei}, \citenamefont {Wang}, \citenamefont {Qiang},
  \citenamefont {Chen}, \citenamefont {Shu},\ and\ \citenamefont
  {Ge}}]{Qiao2021Phys.Rev.C103.034621}%
  \BibitemOpen
  \bibfield  {author} {\bibinfo {author} {\bibfnamefont {C.~Y.}\ \bibnamefont
  {Qiao}}, \bibinfo {author} {\bibfnamefont {J.~C.}\ \bibnamefont {Pei}},
  \bibinfo {author} {\bibfnamefont {Z.~A.}\ \bibnamefont {Wang}}, \bibinfo
  {author} {\bibfnamefont {Y.}~\bibnamefont {Qiang}}, \bibinfo {author}
  {\bibfnamefont {Y.~J.}\ \bibnamefont {Chen}}, \bibinfo {author}
  {\bibfnamefont {N.~C.}\ \bibnamefont {Shu}}, \ and\ \bibinfo {author}
  {\bibfnamefont {Z.~G.}\ \bibnamefont {Ge}},\ }\href {\doibase
  10.1103/physrevc.103.034621} {\bibfield  {journal} {\bibinfo  {journal}
  {Phys. Rev. C}\ }\textbf {\bibinfo {volume} {103}},\ \bibinfo {pages}
  {034621} (\bibinfo {year} {2021})}\BibitemShut {NoStop}%
\bibitem [{\citenamefont {Nik{\v{s}}i{\'{c}}}\ \emph
  {et~al.}(2011)\citenamefont {Nik{\v{s}}i{\'{c}}}, \citenamefont {Vretenar},\
  and\ \citenamefont {Ring}}]{Niksic2011Prog.Part.Nucl.Phys.66.519548}%
  \BibitemOpen
  \bibfield  {author} {\bibinfo {author} {\bibfnamefont {T.}~\bibnamefont
  {Nik{\v{s}}i{\'{c}}}}, \bibinfo {author} {\bibfnamefont {D.}~\bibnamefont
  {Vretenar}}, \ and\ \bibinfo {author} {\bibfnamefont {P.}~\bibnamefont
  {Ring}},\ }\href {\doibase 10.1016/j.ppnp.2011.01.055} {\bibfield  {journal}
  {\bibinfo  {journal} {Prog. Part. Nucl. Phys.}\ }\textbf {\bibinfo {volume}
  {66}},\ \bibinfo {pages} {519} (\bibinfo {year} {2011})}\BibitemShut
  {NoStop}%
\bibitem [{\citenamefont {Liang}\ and\ \citenamefont
  {Litvinova}(2022)}]{Liang2022.}%
  \BibitemOpen
  \bibfield  {author} {\bibinfo {author} {\bibfnamefont {H.}~\bibnamefont
  {Liang}}\ and\ \bibinfo {author} {\bibfnamefont {E.}~\bibnamefont
  {Litvinova}},\ }\href {\doibase 10.48550/ARXIV.2203.02848} {\enquote
  {\bibinfo {title} {Theory of nuclear collective vibrations},}\ } (\bibinfo
  {year} {2022})\BibitemShut {NoStop}%
\bibitem [{\citenamefont {Paar}\ \emph {et~al.}(2007)\citenamefont {Paar},
  \citenamefont {Vretenar}, \citenamefont {Khan},\ and\ \citenamefont
  {Col{\`{o}}}}]{Paar2007Rep.Progr.Phys.70.R02}%
  \BibitemOpen
  \bibfield  {author} {\bibinfo {author} {\bibfnamefont {N.}~\bibnamefont
  {Paar}}, \bibinfo {author} {\bibfnamefont {D.}~\bibnamefont {Vretenar}},
  \bibinfo {author} {\bibfnamefont {E.}~\bibnamefont {Khan}}, \ and\ \bibinfo
  {author} {\bibfnamefont {G.}~\bibnamefont {Col{\`{o}}}},\ }\href {\doibase
  10.1088/0034-4885/70/5/r02} {\bibfield  {journal} {\bibinfo  {journal} {Rep.
  Progr. Phys.}\ }\textbf {\bibinfo {volume} {70}},\ \bibinfo {pages} {R02}
  (\bibinfo {year} {2007})}\BibitemShut {NoStop}%
\bibitem [{\citenamefont {Roca-Maza}\ and\ \citenamefont
  {Paar}(2018)}]{RocaMaza2018Prog.Part.Nucl.Phys.101.96176}%
  \BibitemOpen
  \bibfield  {author} {\bibinfo {author} {\bibfnamefont {X.}~\bibnamefont
  {Roca-Maza}}\ and\ \bibinfo {author} {\bibfnamefont {N.}~\bibnamefont
  {Paar}},\ }\href {\doibase 10.1016/j.ppnp.2018.04.001} {\bibfield  {journal}
  {\bibinfo  {journal} {Prog. Part. Nucl. Phys.}\ }\textbf {\bibinfo {volume}
  {101}},\ \bibinfo {pages} {96} (\bibinfo {year} {2018})}\BibitemShut
  {NoStop}%
\bibitem [{\citenamefont {Zhou}\ \emph {et~al.}(2011)\citenamefont {Zhou},
  \citenamefont {Meng}, \citenamefont {Ring},\ and\ \citenamefont
  {Zhao}}]{Zhou2011J.Phys.Conf.Ser.312.092067}%
  \BibitemOpen
  \bibfield  {author} {\bibinfo {author} {\bibfnamefont {S.~G.}\ \bibnamefont
  {Zhou}}, \bibinfo {author} {\bibfnamefont {J.}~\bibnamefont {Meng}}, \bibinfo
  {author} {\bibfnamefont {P.}~\bibnamefont {Ring}}, \ and\ \bibinfo {author}
  {\bibfnamefont {E.~G.}\ \bibnamefont {Zhao}},\ }\href {\doibase
  10.1088/1742-6596/312/9/092067} {\bibfield  {journal} {\bibinfo  {journal}
  {J. Phys.: Conf. Ser.}\ }\textbf {\bibinfo {volume} {312}},\ \bibinfo {pages}
  {092067} (\bibinfo {year} {2011})}\BibitemShut {NoStop}%
\bibitem [{\citenamefont {Long}\ \emph {et~al.}(2010)\citenamefont {Long},
  \citenamefont {Ring}, \citenamefont {Meng}, \citenamefont {Giai},\ and\
  \citenamefont {Bertulani}}]{Long2010Phys.Rev.C81.031302}%
  \BibitemOpen
  \bibfield  {author} {\bibinfo {author} {\bibfnamefont {W.~H.}\ \bibnamefont
  {Long}}, \bibinfo {author} {\bibfnamefont {P.}~\bibnamefont {Ring}}, \bibinfo
  {author} {\bibfnamefont {J.}~\bibnamefont {Meng}}, \bibinfo {author}
  {\bibfnamefont {N.~V.}\ \bibnamefont {Giai}}, \ and\ \bibinfo {author}
  {\bibfnamefont {C.~A.}\ \bibnamefont {Bertulani}},\ }\href {\doibase
  10.1103/physrevc.81.031302} {\bibfield  {journal} {\bibinfo  {journal} {Phys.
  Rev. C}\ }\textbf {\bibinfo {volume} {81}},\ \bibinfo {pages} {031302}
  (\bibinfo {year} {2010})}\BibitemShut {NoStop}%
\bibitem [{\citenamefont {Moreno}\ \emph {et~al.}(2020)\citenamefont {Moreno},
  \citenamefont {Carleo},\ and\ \citenamefont
  {Georges}}]{Moreno2020Phys.Rev.Lett.125.076402}%
  \BibitemOpen
  \bibfield  {author} {\bibinfo {author} {\bibfnamefont {J.~R.}\ \bibnamefont
  {Moreno}}, \bibinfo {author} {\bibfnamefont {G.}~\bibnamefont {Carleo}}, \
  and\ \bibinfo {author} {\bibfnamefont {A.}~\bibnamefont {Georges}},\ }\href
  {\doibase 10.1103/physrevlett.125.076402} {\bibfield  {journal} {\bibinfo
  {journal} {Phys. Rev. Lett.}\ }\textbf {\bibinfo {volume} {125}},\ \bibinfo
  {pages} {076402} (\bibinfo {year} {2020})}\BibitemShut {NoStop}%
\bibitem [{\citenamefont {Ryczko}\ \emph {et~al.}(2019)\citenamefont {Ryczko},
  \citenamefont {Strubbe},\ and\ \citenamefont
  {Tamblyn}}]{Ryczko2019Phys.Rev.A100.022512}%
  \BibitemOpen
  \bibfield  {author} {\bibinfo {author} {\bibfnamefont {K.}~\bibnamefont
  {Ryczko}}, \bibinfo {author} {\bibfnamefont {D.~A.}\ \bibnamefont {Strubbe}},
  \ and\ \bibinfo {author} {\bibfnamefont {I.}~\bibnamefont {Tamblyn}},\ }\href
  {\doibase 10.1103/physreva.100.022512} {\bibfield  {journal} {\bibinfo
  {journal} {Phys. Rev. A}\ }\textbf {\bibinfo {volume} {100}},\ \bibinfo
  {pages} {022512} (\bibinfo {year} {2019})}\BibitemShut {NoStop}%
\bibitem [{\citenamefont {Kohn}\ and\ \citenamefont
  {Sham}(1965)}]{Kohn1965Phys.Rev.140.A1133A1138}%
  \BibitemOpen
  \bibfield  {author} {\bibinfo {author} {\bibfnamefont {W.}~\bibnamefont
  {Kohn}}\ and\ \bibinfo {author} {\bibfnamefont {L.~J.}\ \bibnamefont
  {Sham}},\ }\href {\doibase 10.1103/physrev.140.a1133} {\bibfield  {journal}
  {\bibinfo  {journal} {Phys. Rev.}\ }\textbf {\bibinfo {volume} {140}},\
  \bibinfo {pages} {A1133} (\bibinfo {year} {1965})}\BibitemShut {NoStop}%
\bibitem [{\citenamefont {Yang}\ \emph {et~al.}(2022)\citenamefont {Yang},
  \citenamefont {Fan}, \citenamefont {Naito}, \citenamefont {Niu},
  \citenamefont {Li},\ and\ \citenamefont {Liang}}]{Yang2022.}%
  \BibitemOpen
  \bibfield  {author} {\bibinfo {author} {\bibfnamefont {Z.-X.}\ \bibnamefont
  {Yang}}, \bibinfo {author} {\bibfnamefont {X.-H.}\ \bibnamefont {Fan}},
  \bibinfo {author} {\bibfnamefont {T.}~\bibnamefont {Naito}}, \bibinfo
  {author} {\bibfnamefont {Z.-M.}\ \bibnamefont {Niu}}, \bibinfo {author}
  {\bibfnamefont {Z.-P.}\ \bibnamefont {Li}}, \ and\ \bibinfo {author}
  {\bibfnamefont {H.}~\bibnamefont {Liang}},\ }\href {\doibase
  10.48550/ARXIV.2205.15649} {\enquote {\bibinfo {title} {Calibration of
  nuclear charge density distribution by back-propagation neural networks},}\ }
  (\bibinfo {year} {2022})\BibitemShut {NoStop}%
\bibitem [{\citenamefont {Bartel}\ \emph {et~al.}(1982)\citenamefont {Bartel},
  \citenamefont {Quentin}, \citenamefont {Brack}, \citenamefont {Guet},\ and\
  \citenamefont {H{\aa}kansson}}]{Bartel1982Nucl.Phys.A386.79100}%
  \BibitemOpen
  \bibfield  {author} {\bibinfo {author} {\bibfnamefont {J.}~\bibnamefont
  {Bartel}}, \bibinfo {author} {\bibfnamefont {P.}~\bibnamefont {Quentin}},
  \bibinfo {author} {\bibfnamefont {M.}~\bibnamefont {Brack}}, \bibinfo
  {author} {\bibfnamefont {C.}~\bibnamefont {Guet}}, \ and\ \bibinfo {author}
  {\bibfnamefont {H.-B.}\ \bibnamefont {H{\aa}kansson}},\ }\href {\doibase
  10.1016/0375-9474(82)90403-1} {\bibfield  {journal} {\bibinfo  {journal}
  {Nucl. Phys. A}\ }\textbf {\bibinfo {volume} {386}},\ \bibinfo {pages} {79}
  (\bibinfo {year} {1982})}\BibitemShut {NoStop}%
\bibitem [{\citenamefont {Yang}\ \emph {et~al.}(2021)\citenamefont {Yang},
  \citenamefont {Fan}, \citenamefont {Yin},\ and\ \citenamefont
  {Zuo}}]{Yang2021Phys.Lett.B823.136650}%
  \BibitemOpen
  \bibfield  {author} {\bibinfo {author} {\bibfnamefont {Z.-X.}\ \bibnamefont
  {Yang}}, \bibinfo {author} {\bibfnamefont {X.-H.}\ \bibnamefont {Fan}},
  \bibinfo {author} {\bibfnamefont {P.}~\bibnamefont {Yin}}, \ and\ \bibinfo
  {author} {\bibfnamefont {W.}~\bibnamefont {Zuo}},\ }\href {\doibase
  10.1016/j.physletb.2021.136650} {\bibfield  {journal} {\bibinfo  {journal}
  {Phys. Lett. B}\ }\textbf {\bibinfo {volume} {823}},\ \bibinfo {pages}
  {136650} (\bibinfo {year} {2021})}\BibitemShut {NoStop}%
\bibitem [{\citenamefont {Ketkar}(2017)}]{Ketkar2017.195208}%
  \BibitemOpen
  \bibfield  {author} {\bibinfo {author} {\bibfnamefont {N.}~\bibnamefont
  {Ketkar}},\ }in\ \href {\doibase 10.1007/978-1-4842-2766-4_12} {\emph
  {\bibinfo {booktitle} {Deep Learning with Python}}}\ (\bibinfo  {publisher}
  {Apress},\ \bibinfo {year} {2017})\ pp.\ \bibinfo {pages}
  {195--208}\BibitemShut {NoStop}%
\bibitem [{\citenamefont {Kingma}\ and\ \citenamefont
  {Ba}(2015)}]{Kingma2015.}%
  \BibitemOpen
  \bibfield  {author} {\bibinfo {author} {\bibfnamefont {D.~P.}\ \bibnamefont
  {Kingma}}\ and\ \bibinfo {author} {\bibfnamefont {J.}~\bibnamefont {Ba}},\
  }in\ \href {http://arxiv.org/abs/1412.6980} {\emph {\bibinfo {booktitle} {3rd
  International Conference on Learning Representations, {ICLR} 2015, San Diego,
  CA, USA, May 7-9, 2015, Conference Track Proceedings}}},\ \bibinfo {editor}
  {edited by\ \bibinfo {editor} {\bibfnamefont {Y.}~\bibnamefont {Bengio}}\
  and\ \bibinfo {editor} {\bibfnamefont {Y.}~\bibnamefont {LeCun}}}\ (\bibinfo
  {year} {2015})\BibitemShut {NoStop}%
\bibitem [{\citenamefont {Sugahara}\ and\ \citenamefont
  {Toki}(1994)}]{Sugahara1994Nucl.Phys.A579.557572}%
  \BibitemOpen
  \bibfield  {author} {\bibinfo {author} {\bibfnamefont {Y.}~\bibnamefont
  {Sugahara}}\ and\ \bibinfo {author} {\bibfnamefont {H.}~\bibnamefont
  {Toki}},\ }\href {\doibase 10.1016/0375-9474(94)90923-7} {\bibfield
  {journal} {\bibinfo  {journal} {Nucl. Phys. A}\ }\textbf {\bibinfo {volume}
  {579}},\ \bibinfo {pages} {557} (\bibinfo {year} {1994})}\BibitemShut
  {NoStop}%
\bibitem [{\citenamefont {Reinhard}(1991)}]{Reinhard1991.2850}%
  \BibitemOpen
  \bibfield  {author} {\bibinfo {author} {\bibfnamefont {P.-G.}\ \bibnamefont
  {Reinhard}},\ }in\ \href {\doibase 10.1007/978-3-642-76356-4_2} {\emph
  {\bibinfo {booktitle} {Computational Nuclear Physics 1}}}\ (\bibinfo
  {publisher} {Springer Berlin Heidelberg},\ \bibinfo {year} {1991})\ pp.\
  \bibinfo {pages} {28--50}\BibitemShut {NoStop}%
\bibitem [{\citenamefont {Xia}\ \emph {et~al.}(2018)\citenamefont {Xia},
  \citenamefont {Lim}, \citenamefont {Zhao}, \citenamefont {Liang},
  \citenamefont {Qu}, \citenamefont {Chen}, \citenamefont {Liu}, \citenamefont
  {Zhang}, \citenamefont {Zhang}, \citenamefont {Kim},\ and\ \citenamefont
  {Meng}}]{Xia2018At.DataNucl.DataTables121122.1215}%
  \BibitemOpen
  \bibfield  {author} {\bibinfo {author} {\bibfnamefont {X.}~\bibnamefont
  {Xia}}, \bibinfo {author} {\bibfnamefont {Y.}~\bibnamefont {Lim}}, \bibinfo
  {author} {\bibfnamefont {P.}~\bibnamefont {Zhao}}, \bibinfo {author}
  {\bibfnamefont {H.}~\bibnamefont {Liang}}, \bibinfo {author} {\bibfnamefont
  {X.}~\bibnamefont {Qu}}, \bibinfo {author} {\bibfnamefont {Y.}~\bibnamefont
  {Chen}}, \bibinfo {author} {\bibfnamefont {H.}~\bibnamefont {Liu}}, \bibinfo
  {author} {\bibfnamefont {L.}~\bibnamefont {Zhang}}, \bibinfo {author}
  {\bibfnamefont {S.}~\bibnamefont {Zhang}}, \bibinfo {author} {\bibfnamefont
  {Y.}~\bibnamefont {Kim}}, \ and\ \bibinfo {author} {\bibfnamefont
  {J.}~\bibnamefont {Meng}},\ }\href {\doibase 10.1016/j.adt.2017.09.001}
  {\bibfield  {journal} {\bibinfo  {journal} {At. Data Nucl. Data Tables}\
  }\textbf {\bibinfo {volume} {121-122}},\ \bibinfo {pages} {1} (\bibinfo
  {year} {2018})}\BibitemShut {NoStop}%
\end{thebibliography}%


\begin{thebibliography}{4}%
\makeatletter
\providecommand \@ifxundefined [1]{%
 \@ifx{#1\undefined}
}%
\providecommand \@ifnum [1]{%
 \ifnum #1\expandafter \@firstoftwo
 \else \expandafter \@secondoftwo
 \fi
}%
\providecommand \@ifx [1]{%
 \ifx #1\expandafter \@firstoftwo
 \else \expandafter \@secondoftwo
 \fi
}%
\providecommand \natexlab [1]{#1}%
\providecommand \enquote  [1]{``#1''}%
\providecommand \bibnamefont  [1]{#1}%
\providecommand \bibfnamefont [1]{#1}%
\providecommand \citenamefont [1]{#1}%
\providecommand \href@noop [0]{\@secondoftwo}%
\providecommand \href [0]{\begingroup \@sanitize@url \@href}%
\providecommand \@href[1]{\@@startlink{#1}\@@href}%
\providecommand \@@href[1]{\endgroup#1\@@endlink}%
\providecommand \@sanitize@url [0]{\catcode `\\12\catcode `\$12\catcode
  `\&12\catcode `\#12\catcode `\^12\catcode `\_12\catcode `\%12\relax}%
\providecommand \@@startlink[1]{}%
\providecommand \@@endlink[0]{}%
\providecommand \url  [0]{\begingroup\@sanitize@url \@url }%
\providecommand \@url [1]{\endgroup\@href {#1}{\urlprefix }}%
\providecommand \urlprefix  [0]{URL }%
\providecommand \Eprint [0]{\href }%
\providecommand \doibase [0]{http://dx.doi.org/}%
\providecommand \selectlanguage [0]{\@gobble}%
\providecommand \bibinfo  [0]{\@secondoftwo}%
\providecommand \bibfield  [0]{\@secondoftwo}%
\providecommand \translation [1]{[#1]}%
\providecommand \BibitemOpen [0]{}%
\providecommand \bibitemStop [0]{}%
\providecommand \bibitemNoStop [0]{.\EOS\space}%
\providecommand \EOS [0]{\spacefactor3000\relax}%
\providecommand \BibitemShut  [1]{\csname bibitem#1\endcsname}%
\let\auto@bib@innerbib\@empty
\bibitem [{\citenamefont {Kingma}\ and\ \citenamefont
  {Ba}(2015)}]{Kingma2015.}%
  \BibitemOpen
  \bibfield  {author} {\bibinfo {author} {\bibfnamefont {D.~P.}\ \bibnamefont
  {Kingma}}\ and\ \bibinfo {author} {\bibfnamefont {J.}~\bibnamefont {Ba}},\
  }in\ \href {http://arxiv.org/abs/1412.6980} {\emph {\bibinfo {booktitle} {3rd
  International Conference on Learning Representations, {ICLR} 2015, San Diego,
  CA, USA, May 7-9, 2015, Conference Track Proceedings}}},\ \bibinfo {editor}
  {edited by\ \bibinfo {editor} {\bibfnamefont {Y.}~\bibnamefont {Bengio}}\
  and\ \bibinfo {editor} {\bibfnamefont {Y.}~\bibnamefont {LeCun}}}\ (\bibinfo
  {year} {2015})\BibitemShut {NoStop}%
\bibitem [{\citenamefont {Reinhard}(1991)}]{Reinhard1991.2850}%
  \BibitemOpen
  \bibfield  {author} {\bibinfo {author} {\bibfnamefont {P.-G.}\ \bibnamefont
  {Reinhard}},\ }in\ \href {\doibase 10.1007/978-3-642-76356-4_2} {\emph
  {\bibinfo {booktitle} {Computational Nuclear Physics 1}}}\ (\bibinfo
  {publisher} {Springer Berlin Heidelberg},\ \bibinfo {year} {1991})\ pp.\
  \bibinfo {pages} {28--50}\BibitemShut {NoStop}%
\bibitem [{\citenamefont {Bartel}\ \emph {et~al.}(1982)\citenamefont {Bartel},
  \citenamefont {Quentin}, \citenamefont {Brack}, \citenamefont {Guet},\ and\
  \citenamefont {H{\aa}kansson}}]{Bartel1982Nucl.Phys.A386.79100}%
  \BibitemOpen
  \bibfield  {author} {\bibinfo {author} {\bibfnamefont {J.}~\bibnamefont
  {Bartel}}, \bibinfo {author} {\bibfnamefont {P.}~\bibnamefont {Quentin}},
  \bibinfo {author} {\bibfnamefont {M.}~\bibnamefont {Brack}}, \bibinfo
  {author} {\bibfnamefont {C.}~\bibnamefont {Guet}}, \ and\ \bibinfo {author}
  {\bibfnamefont {H.-B.}\ \bibnamefont {H{\aa}kansson}},\ }\href {\doibase
  10.1016/0375-9474(82)90403-1} {\bibfield  {journal} {\bibinfo  {journal}
  {Nucl. Phys. A}\ }\textbf {\bibinfo {volume} {386}},\ \bibinfo {pages} {79}
  (\bibinfo {year} {1982})}\BibitemShut {NoStop}%
\bibitem [{\citenamefont {Yang}\ \emph {et~al.}(2021)\citenamefont {Yang},
  \citenamefont {Fan}, \citenamefont {Yin},\ and\ \citenamefont
  {Zuo}}]{Yang2021Phys.Lett.B823.136650}%
  \BibitemOpen
  \bibfield  {author} {\bibinfo {author} {\bibfnamefont {Z.-X.}\ \bibnamefont
  {Yang}}, \bibinfo {author} {\bibfnamefont {X.-H.}\ \bibnamefont {Fan}},
  \bibinfo {author} {\bibfnamefont {P.}~\bibnamefont {Yin}}, \ and\ \bibinfo
  {author} {\bibfnamefont {W.}~\bibnamefont {Zuo}},\ }\href {\doibase
  10.1016/j.physletb.2021.136650} {\bibfield  {journal} {\bibinfo  {journal}
  {Phys. Lett. B}\ }\textbf {\bibinfo {volume} {823}},\ \bibinfo {pages}
  {136650} (\bibinfo {year} {2021})}\BibitemShut {NoStop}%
\end{thebibliography}%

\end{document}


\beginsupplement
\preprint{RIKEN-iTHEMS-Report-22}

\begin{CJK*}{UTF8}{gbsn}

\title{Supplemental Materials for ``A Kohn-Sham Scheme Based Neural Network for Nuclear Systems"}

\author{Zu-Xing Yang}
\affiliation{RIKEN Nishina Center, Wako 351-0198, Japan}

\author{Xiao-Hua Fan}
\affiliation{School of Physical Science and Technology, Southwest University, Chongqing 400715, China}

\author{Zhi-Pan Li}
\affiliation{School of Physical Science and Technology, Southwest University, Chongqing 400715, China}

\author{Haozhao Liang}
\affiliation{Department of Physics, Graduate School of Science, The University of Tokyo, Tokyo 113-0033, Japan}
\affiliation{RIKEN Interdisciplinary Theoretical and Mathematical Sciences Program, Wako 351-0198, Japan}


\maketitle
\end{CJK*}

\section{details of Neural network structure and Machine Learning Process}

In addition to Fig.~1, the 30 FC neural network branch cells with the same structure ($\mathrm{C2}$) correspond to the 30 single-particle states, where the typical shell ordering according to the nuclear oscillator shell model is taken for the state $S_i$ listed in Table~\ref{tb_shellorder}.
Even though the ordering of states within two shell closures may differ from model to model, the objective functions guide the network to learn the correct sequence.

\begin{table}[h]
\caption{Typical shell ordering according to the nuclear oscillator shell model for protons as well as for neutrons. The ``degeneracy" is the maximum number of nucleons in the given single-particle state.}
\begin{tabular}{ccc|ccc|ccc}
\hline
\hline
order ($i$) & name ($S_i$)  & degeneracy ($d_i$) & order ($i$) & name ($S_i$) & degeneracy ($d_i$) & order ($i$) & name ($S_i$)  & degeneracy ($d_i$) \\ \hline
1     & $1s_{1/2}$ & 2          & 11    & $1g_{9/2}$  & 10         & 21    & $3p_{1/2}$  & 2          \\
2     & $1p_{3/2}$ & 4          & 12    & $1g_{7/2}$  & 8          & 22    & $1i_{13/2}$& 14         \\
3     & $1p_{1/2}$ & 2          & 13    & $2d_{5/2}$  & 6          & 23    & $1i_{11/2}$ & 12         \\
4     & $1d_{5/2}$ & 6          & 14    & $2d_{3/2}$  & 4          & 24    & $2g_{9/2}$  & 10         \\
5     & $1d_{3/2}$ & 4          & 15    & $3s_{1/2}$ & 2          & 25    & $2g_{7/2}$  & 8          \\
6     & $2s_{1/2}$ & 2          & 16    & $1h_{11/2}$ & 12         & 26    & $4d_{5/2}$ & 6          \\
7     & $1f_{7/2}$ & 8          & 17    & $1h_{9/2}$  & 10         & 27    & $4d_{3/2}$  & 4          \\
8     & $1f_{5/2}$& 6          & 18    & $2f_{7/2}$  & 8          & 28    & $5s_{1/2}$ & 2          \\
9     & $2p_{3/2}$ & 4          & 19    & $3f_{5/2}$  & 6          & 29    & $1j_{15/2}$ & 16         \\
10    & $1p_{1/2}$ & 2          & 20    & $3p_{3/2}$  & 4          & 30    & $1j_{13/2}$ & 14         \\ \hline\hline
\end{tabular}
\label{tb_shellorder}
\end{table}

For the normalization module, considering the conservation of particle number and the normalized single-particle wave functions, the excess nucleon number $\Delta$ is superimposed on the outermost shell according to typical shell ordering denoted by
\begin{equation}
\sqrt{w_i} \varphi_i(r) = \hat{N}(\sqrt{w'_i} \varphi_i(r)).
\label{eq_nOR}
\end{equation}
In implementation, the normalization operator is achieved via the following three steps.
\begin{enumerate}
\item  Extracting and activating $w'_i$:

The $\varphi_{i}$ is considered as normalized so that occupancy probability $w'_i$ can be extracted by
\begin{equation}
\label{eq_w}
w'_i = \int (\sqrt{w'_i} \varphi_i(r))^2 r^2 dr, 
\end{equation}
whose value domain is truncated to be $[0,1]$ by a novel activation function
\begin{equation}
w''_i = g(w'_i) =\begin{cases}1 & w'_i > 1 \\ w'_i & w'_i \in [0,1] \end{cases}.
\end{equation}

\item Calculating the excess nucleon number $\Delta$ :

Due to the limitations of the activation function $\mathrm{tanh}(x)$ in C2 and the pre-trained process, the excess nucleon number $\Delta$ would be a small value noted as
\begin{equation}
\Delta = A^\tau - \sum_i w''_i d_i.
\end{equation}

\item Calibrating occupancy probabilities:

As a simple treatment, the $\Delta$ is added to the outermost shell $I$ according to typical shell ordering (see Table~\ref{tb_shellorder}), i.e., $I$ satisfies
\begin{equation}
\sum_{i=1}^{I-1} d_i < A^\tau < \sum_{i=1}^{I} d_i,
\end{equation}
and the final $w_i$ that strictly obeys the conservation of particle number is denoted as
\begin{equation}
\label{eq_10}
w_i =\begin{cases}w''_i + \Delta/d_i & i=I   \\ w''_i & i\neq I \end{cases}.
\end{equation}

\end{enumerate}

\begin{table}[h]
\caption{ The hyperparameters sets of two FC neural network cells.  The $S_l$ is number of neurons and the $g(x)$ is activation function.}
\begin{tabular}{llll}
\hline\hline
\multicolumn{4}{l}{Trunk cell (C1)}                                               \\ \hline
The $l$-th layer~~~ & Name~~~~~~~~     & $S_l$~~~~ & $g(x)$~~~~ \\
0              & Input    & 2                 & ---                  \\
1              & FC Layer & 64                & ReLU                 \\
2              & FC Layer & 128               & ReLU                 \\
3              & FC Layer & 256               & ReLU                 \\
4              & FC Layer & 512               & ReLU                 \\
5              & FC Layer & 1024              & ReLU                 \\ \hline
\multicolumn{4}{l}{Branch cell (C2)}                                               \\ \hline
The $l$-th layer & Name     & $S_l$ & $g(x)$ \\
1              & FC Layer & 512               & ReLU                 \\
2              & FC Layer & 256               & ReLU                 \\
3              & FC Layer & 150               & tanh                 \\ \hline
\multicolumn{4}{l}{Other information}   \\
\hline
\multicolumn{2}{l}{Objective function (1-1500)} & \multicolumn{2}{l}{$\mathrm{Loss}_1$} \\
\multicolumn{2}{l}{Objective function (1501-2000)} & \multicolumn{2}{l}{$\mathrm{Loss}_2$} \\
\multicolumn{2}{l}{Optimizer} & \multicolumn{2}{l}{Adam \cite{Kingma2015.}} \\
\multicolumn{2}{l}{$lr$ (1-500)} & \multicolumn{2}{l}{$5 \times 10^{-4}$} \\
\multicolumn{2}{l}{$lr$ (501-1000)} & \multicolumn{2}{l}{$1 \times 10^{-4}$} \\
\multicolumn{2}{l}{$lr$ (1001-1700)} & \multicolumn{2}{l}{$1 \times 10^{-5}$} \\
\multicolumn{2}{l}{$lr$ (1701-1800)} & \multicolumn{2}{l}{$1 \times 10^{-6}$} \\
\multicolumn{2}{l}{$lr$ (1801-2000)} & \multicolumn{2}{l}{$1 \times 10^{-7}$} \\
\hline \hline
\end{tabular}
\label{tb_hyp}
\end{table}

To feasibly operationalize this training, we have to decompose the whole process into the two parts and vary the learning rate ($lr$) according to the loss value.
The carefully modulated hyperparameter sets of the fully connected (FC) neural network cells and the varying learning rate are listed in Table~\ref{tb_hyp}.

\begin{figure}[htb]
\includegraphics[width=14 cm]{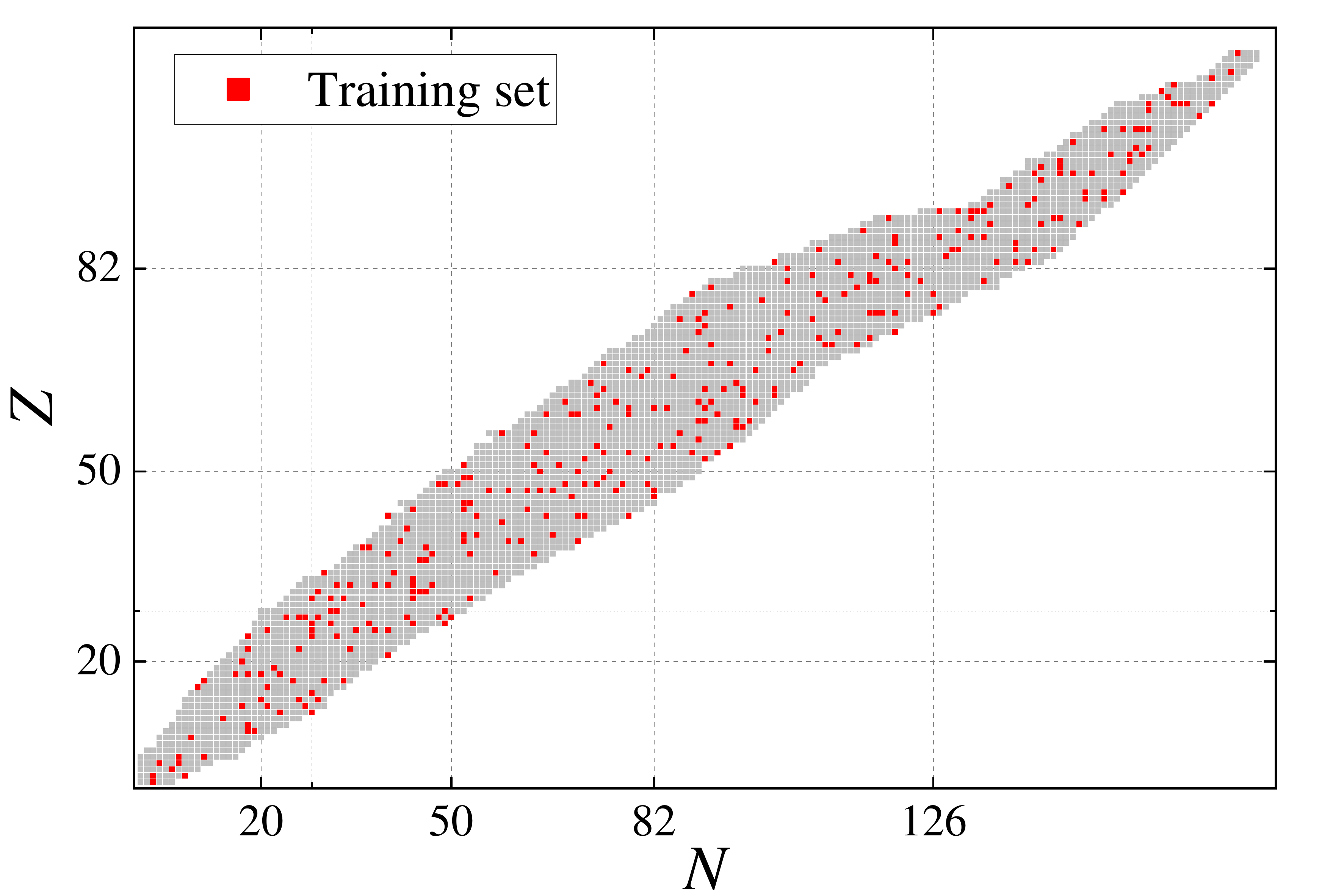}
\caption{\label{fig:trainset} (Color online) The training set including 320 trained nuclei.
The red symbols mark the trained nuclei and the grey symbols indicate the other nuclei discovered in laboratories.
}
\end{figure}

In application, the random 320 nuclei shown in Fig.~\ref{fig:trainset} are adopted to train Kohn-Sham network (KSN), where the trained nuclei are marked by red symbols and the other nuclei discovered in laboratories are indicated by grey symbols.

\section{detailed information in theoretical calculations}

In this section, the detailed parameters of theoretical calculations are presented in Table \ref{tb_theopara}. 
We employed the program HAFOMN \cite{Reinhard1991.2850} to calculate the nuclear ground-state wave functions in the Skyrme Hartree-Fock  framework including the Bardeen-Cooper-Schrieffer pairing (SHF+BCS).
The parameters in the table correspond to the Skyrme density functional we used, SkM* \cite{Bartel1982Nucl.Phys.A386.79100}.
All other parameters not listed are set to their default values in the program.

\begin{table}[htb]
\caption{The parameters of program HAFOMN. See Ref.~\cite{Reinhard1991.2850} for the meanings of the abbreviations.}
\begin{tabular}{ll|ll|ll|ll}
\hline\hline
~~~Property~~~ & Value~~~  & ~~~Property~~~ & Value~~~ & ~~~Property~~~ & Value~~~ & ~~~Property~~~  & Value~~~   \\ \hline
~~~RSTEP    & .04   & ~~~XODAMP   & 0     & ~~~IFEX     & 1     & ~~~IFDFMS    & 0       \\
~~~ITMAX    & 100    & ~~~EODAMP   & 40    & ~~~IFTM     & 0     & ~~~IFRHOC    & 0       \\
~~~EPSPR    & .00001 & ~~~ADDNEW   & 5     & ~~~IFZPE    & 0     & ~~~PAIRFP(N) & -1.1E11 \\ \hline\hline
\end{tabular}
\label{tb_theopara}
\end{table}

\section{Extrapolation results for Pb isotopes}

\begin{figure}[htb]
\includegraphics[width=18 cm]{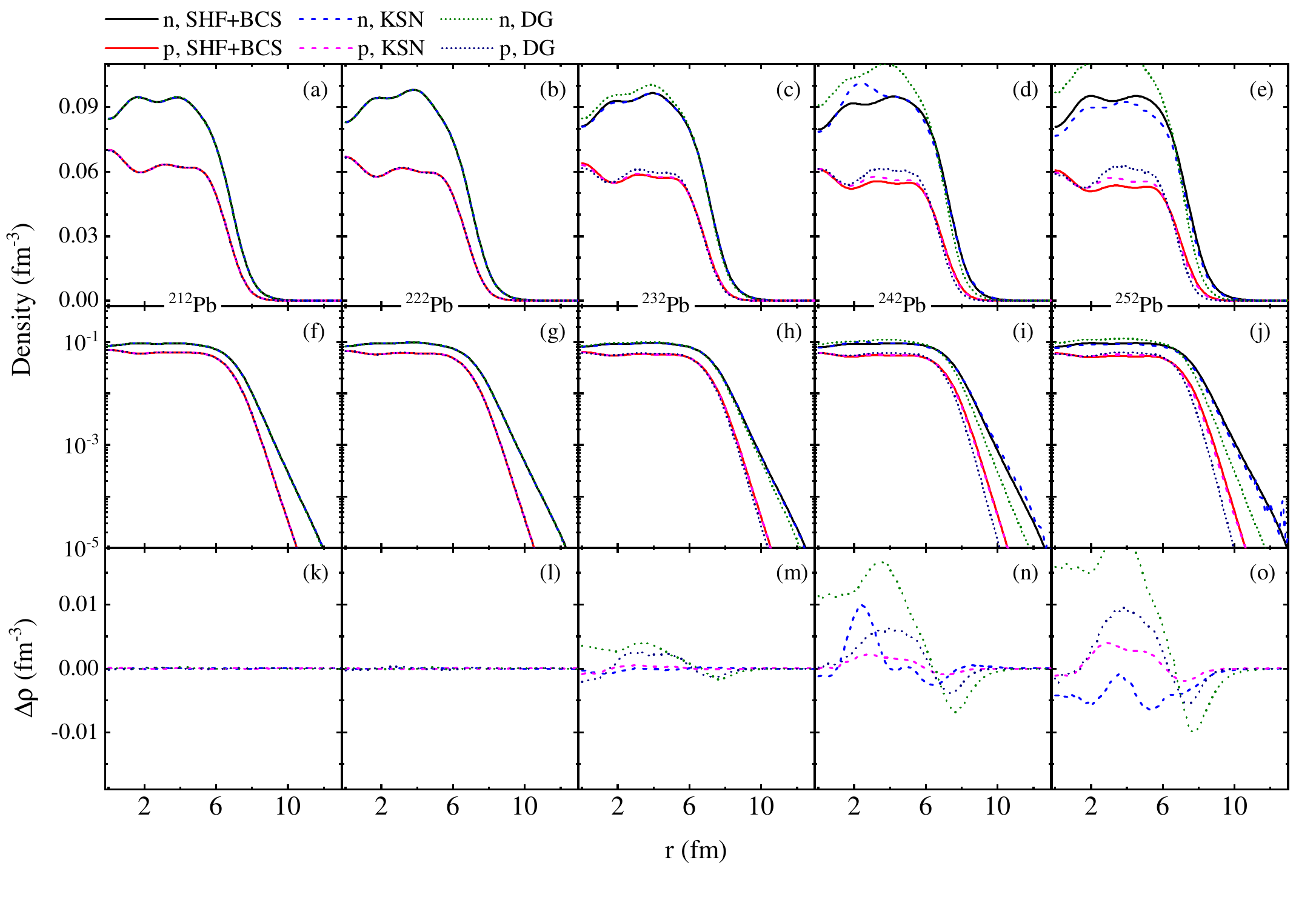}
\caption{\label{fig:extrapden} (Color online) Upper and middle panels: Neutron and proton densities predicted by KSN with 2400 trained nuclei for the nuclei $^{212}$Pb, $^{222}$Pb, $^{232}$Pb, $^{242}$Pb, and $^{252}$Pb in the linear and logarithmic scales. 
The SHF+BCS results and DG results are also shown for comparison.
Lower panels: The corresponding deviations $\Delta \rho$ between the predictions by neural networks (KSN/DG) and the calculations by SHF+BCS.
}
\end{figure}

\begin{figure}[htb]
\includegraphics[width=13 cm]{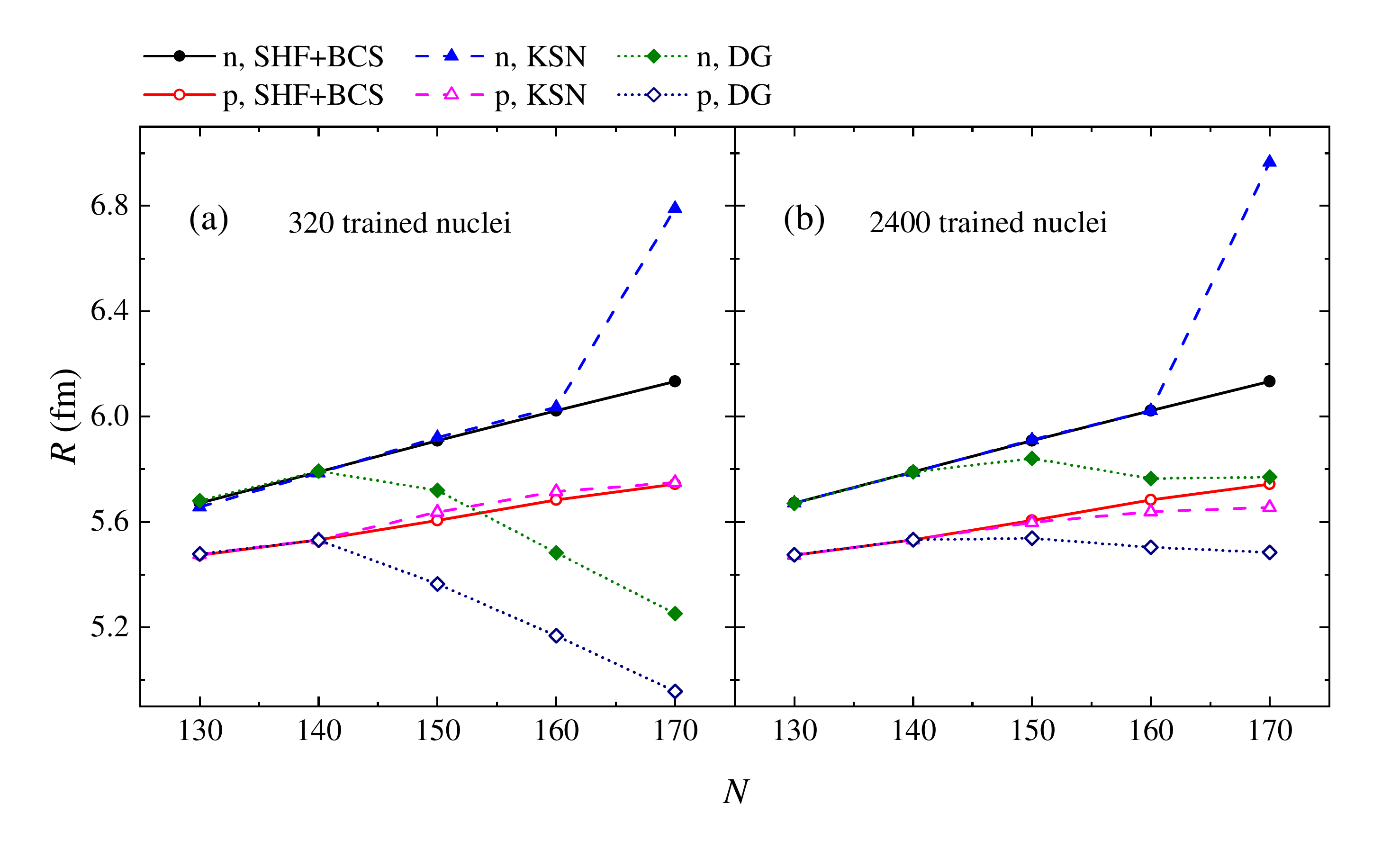}
\caption{\label{fig:DR} (Color online) Neutron and proton radii $R$ for the Pb isotopes predicted by KSN and DG with (a) 320 nuclei trained and (b) 2400 nuclei trained.}
\end{figure}

In this section, in addition to Fig.~4, we select randomly the 2400 nuclei found in laboratories as training and extrapolate farther for Pb. 
Figure~\ref{fig:extrapden} (a)-(j) show the nuclear densities predicted by KSN and density generator (DG) \cite{Yang2021Phys.Lett.B823.136650} for the nuclei $^{212}$Pb, $^{222}$Pb, $^{232}$Pb, $^{242}$Pb, and $^{252}$Pb in linear and logarithmic scales.
Meanwhile, the corresponding deviations $\Delta \rho$ between the predictions by neural networks (KSN/DG) and the calculations by SHF+BCS are presented in (k)-(o).
Figure~\ref{fig:extrapden} displays that an obvious deviation appears in $^{232}$Pb for DG, while for KSN, no deviation appears until $^{242}$Pb.
Consistent with Fig.~4, the predictions of KSN are much better than those of DG amid the increasing deviation with extrapolation, which again proves the key conclusion in the main text, i.e., neural networks extrapolation naturally improves as more physical properties are learned and more constraints are imposed.

As a comparison, the nuclear radii $R$ for the Pb isotopes predicted by KSN and DG are shown in Fig.~\ref{fig:DR} with (a) 320 nuclei trained and (b) 2400 nuclei trained.
Remarkably, the extrapolations of KSN in nuclear radii are more steady and effective than those of DG, where the KSN deviates not large until $^{242}$Pb but DG fails in the case of $^{232}$Pb.
As the shell evolution is achieved, KSN obtains more accurate predictions based on less data, which clearly illustrates the importance of physics in a data-driven approach.

\clearpage

\bibliographystyle{apsrev4-1}
\bibliography{ref}